\documentclass[floatfix, twocolumn,amsmath,amssymb, aps,prb,superscriptaddress]{revtex4-2}
\usepackage{mathrsfs}
\usepackage{amsmath}
\usepackage{ragged2e}

\usepackage{caption}
\usepackage{subcaption}
\usepackage{graphicx}
\usepackage{hyperref}
\usepackage{stfloats}
\usepackage{ulem}
\hypersetup{ colorlinks, citecolor=blue, linkcolor=blue, urlcolor=blue}

\DeclareCaptionLabelFormat{nocaption}{}
\graphicspath{{img/}, {img/img-pdbands/}}

\begin{document}
	\title{Band engineered bilayer Haldane model: Evidence of multiple topological phase transitions}
	
	\author{Sayan Mondal}
	\affiliation{Department of Physics, Indian Institute of Technology Guwahati, 
		Guwahati 781039, Assam, India}
	\author{Saurabh Basu}
	\affiliation{Department of Physics, Indian Institute of Technology Guwahati, 
		Guwahati 781039, Assam, India}

	\begin{abstract}
	
	We have studied the evolution of the topological properties of a band-engineered AB-stacked bilayer honeycomb structure in the presence of a Haldane flux. Without a Haldane flux, band engineering makes the band touching points (the so called Dirac points) move towards each other and eventually merge into one at an intermediate $\mathbf{M}$ point in the Brillouin zone. Here the dispersion is linear along one direction and quadratic along the other. In the presence of a Haldane flux, the system acquires topological properties, and finite Chern numbers can be associated with the pairs of the conduction and the valence bands. The valence band closer to the Fermi level ($E_F$) possesses Chern numbers equal to $\pm2$ and $\pm1$, while the one further away from $E_F$ corresponds to Chern numbers $\pm1$. The conduction bands are associated with similar properties, except their signs are reversed. The Chern lobes shrink in the band-engineered model, and we find evidence of multiple topological phase transitions, where the Chern numbers discontinuously jump from $\pm2$ to $\mp2$, $\pm1$ to $\mp1$, $\pm1$ to $0$ to $\pm2$ and $\pm2$ to $\pm1$. These transitions are supported by the presence or absence of the chiral edge modes in a nanoribbon bilayer geometry and the vanishing of the plateau in the anomalous Hall conductivity. Different phases are further computed for different hopping amplitudes across the layers, which shows shrinking of the Chern lobes for large interlayer tunneling.

	\end{abstract}
	\maketitle
	\section{Introduction}\label{sec:intro}
	
	The Haldane model is a toy model, where it was shown that one can achieve quantum Hall effect even in the absence of an external magnetic field in a two-dimensional honeycomb lattice \cite{haldane1988}. To achieve such a scenario, the time-reversal symmetry (TRS) of the system needs to be broken which can be done via chiral complex next nearest neighbour hopping amplitudes. The spectral bands of such a system possess a non-zero topological invariant known as the Chern number, and hence the system is known as a Chern insulator. Futhermore, the band structure of a semi-infinite ribbon geometry hosts chiral edge modes which is the signature of its topological character. Also, the system exhibits quantum anomalous Hall effect which shows a plateau structure in the vicinity of zero Fermi energy. 
	
	Haldane's work has trigged an extensive study in both theoretical and experimental fronts. for example, there have been reports of Haldane-like spectrum and non-trivial phases realized in dice lattice \cite{kapri2020}, Kagom\'e lattice \cite{ohgushi2000, xiao2003, guo2009, liu2013}, checkerboard lattice \cite{sun2009}, Lieb lattice \cite{weeks2010, apaja2010, goldman2011, tsai2015}, buckled lattice \cite{wright2013} etc. Experimentally, Haldane model has been realized in cold atoms situated at the optical lattices where the complex second neighbour hopping can be created by means of standing-wave laser beams \cite{shao2008, alba2011, tarruel2012}, ultracold fermions in the optical honeycomb lattices \cite{jotzu2014} etc. Also, a two-dimensional honeycomb structure of Fe-based insulators, such as, $X\mathrm{Fe_2(PO_4)_2}$, with $X$ being either one of these, K, Cs, La \cite{kim2017} demonstrate similar non-trivial topological phases with a non-zero Chern number. Further, non-zero Chern numbers have also been found in acoustic Chern insulators \cite{ding2019}, the interface between the two trivial ferromagnetic insulators EuO and GdN \cite{vanderbilt2014} etc.
	
	In recent years, there have been studies of the Haldane model in coupled two-dimensional systems, for example, the bilayer materials \cite{spurrier2020,cheng2019,sorn2018,panas2020}. In parallel, there are studies on the band engineering in various systems, such as, single layer graphene \cite{mondal2021}, spin Hall insulators \cite{mondal2022_1} and a dice lattice \cite{mondal2023_1}. Such a band engineering has been incorporated via the introduction of an anisotropy among the nearest neighbour (NN) hoppings. Such hopping anisotropies have been included between the neighbouring sites lying along a particular direction (say, $t_1$), while keeping the rest NN hoppings as $t$ in a honeycomb lattice. If the value of $t_1$ is varied, the band extrema from the two Dirac points move closer to each other and they finally merge with a vanishing band gap at the $\mathbf{M}$ point in the Brillouin zone (BZ) for a particular value of $t_1$, namely, $t_1 = 2t$, which is called as the semi-Dirac limit. During the process, the topological properties of the system also vanishes at the gap closing hopping amplitude $t_1 = 2t$. It should be noted that the band structure of the system in absence of the complex NNN hopping (Haldane flux) shows semi-Dirac dispersion, that is, linear along the $k_x$-direction and quadratic along the $k_y$-direction. Experimentally, the semi-Dirac dispersion has been observed in many materials, such as, multilayered structures of $\mathrm{TiO}_2/\mathrm{VO}_2$ \cite{pickett2009,pickett2010}, monolayer phosphorene in presence of doping and pressure \cite{rodin2014,guan2014}, $\mathrm{BEDT}$-$\mathrm{TTF_2I_3}$ organic salts under pressure \cite{suzumura2013,hasegawa2006}, black phosphorus doped with potassium atoms by means of \textit{in situ} deposition \cite{kim2015} etc. One can also achieve the semi-Dirac dispersion by applying an uniaxial strain to a system which will change the bond length lying along a particular direction which is parallel to the applied strain direction. Therefore, the hopping energies along those directions get modified, while the hopping along the other directions remains unaltered. Such method has been employed in a monolayer honeycomb structure, such as, $\mathrm{Si_2O}$ which yields a semi-Dirac dispersion \cite{zhong2017}. 
	
	However, the effect of band engineering in a multi-layered system, such as, a bilayer graphene has never been studied. Needless to mention that bilayers possess a richer phase diagram comprising of a larger parameter space. The topological properties of such an engineered system are interesting since the existence of the edge modes and the quantized Hall conductivity have never been studied. A more interesting issue is that owing to larger number of bands being present in the band structure of a bilayer system, higher values of Chern numbers are realized. A higher Chern number implies higher value of the anomalous Hall conductivity together with larger number of chiral edge modes present in a semi-infinite system. 
	
	Higher Chern numbers are in general interesting and can be realized in a host of systems, such as, in the Dirac \cite{sticlet2013} and semi-Dirac \cite{mondal2022_2} systems in presence of longer range hopping, multi-orbital triangular lattices \cite{sarma2012}, star lattices or decorated honeycomb lattices \cite{chen2012}, honeycomb lattices in presence of spin-orbit coupling \cite{yang2014, yang2016}, ultracold gases in triangular lattices \cite{alase2021, goldman2021} etc. Further, topological insulators doped with magnetic materials \cite{bernevig2014}, Cr-doped thin laminar sheets of $\mathrm{Bi}_2(\mathrm{Se, Te})_3$ \cite{zhang2013} also demonstrate higher values of the Chern numbers. Further, $\mathrm{MnBi}_2\mathrm{Te}_4$ at high temperature \cite{ge2020, zhu2022}, multilayered structure of doped (with magnetic materials) and undoped topological insulators arranged alternatively \cite{samarth2020}, and in classical systems, such as, sonic crystals prepared using acoustic components \cite{zhao2022} show non-trivial phases with higher Chern numbers.
	
		\begin{figure}[h]
		\centering
		\begin{subfigure}[b]{0.8\linewidth}
			\includegraphics[width=\textwidth]{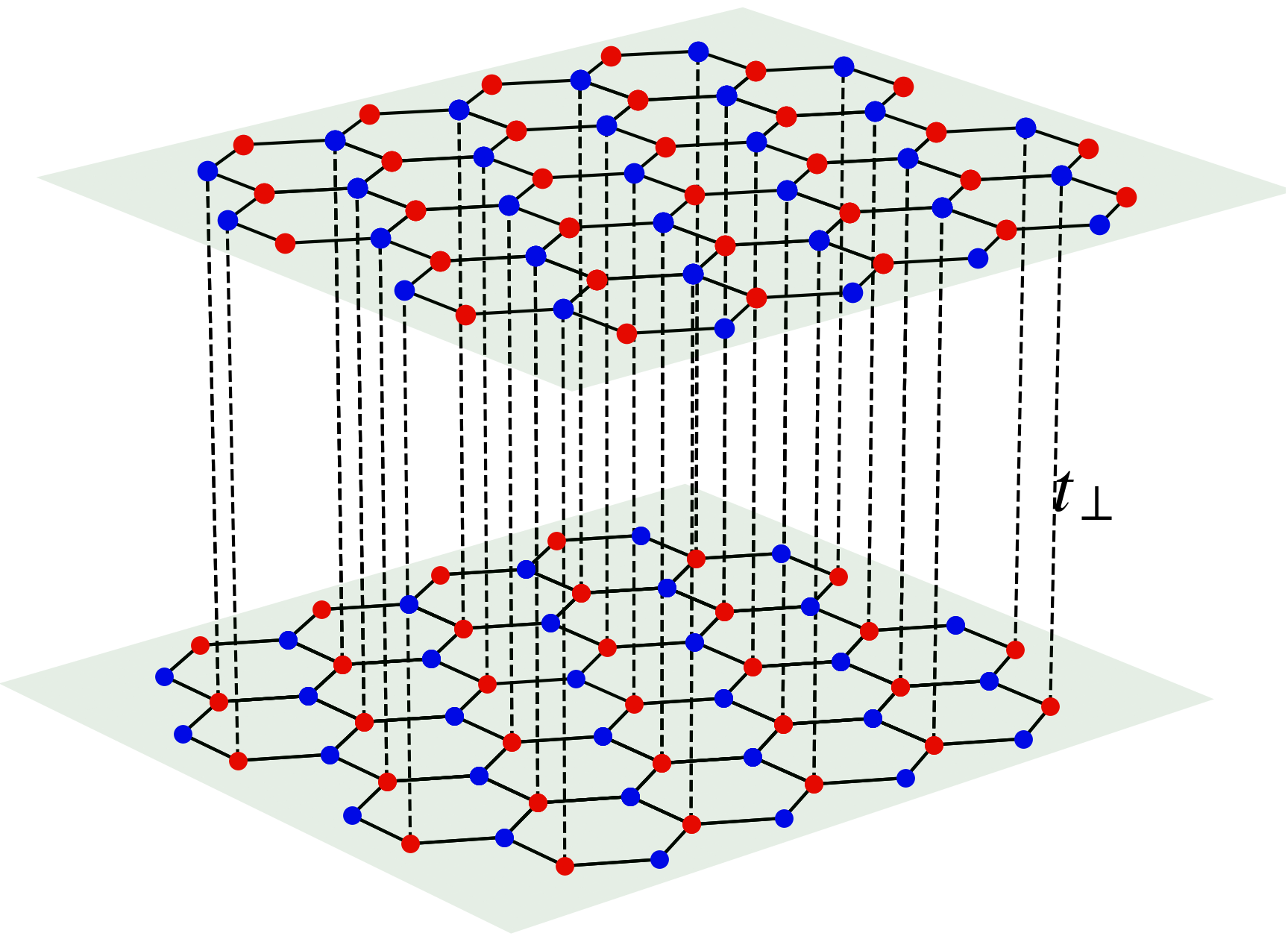}
			\subcaption{}\label{fig:bilayer_stacking}
		\end{subfigure}
		\begin{subfigure}[b]{0.8\linewidth}
			\includegraphics[width=\textwidth]{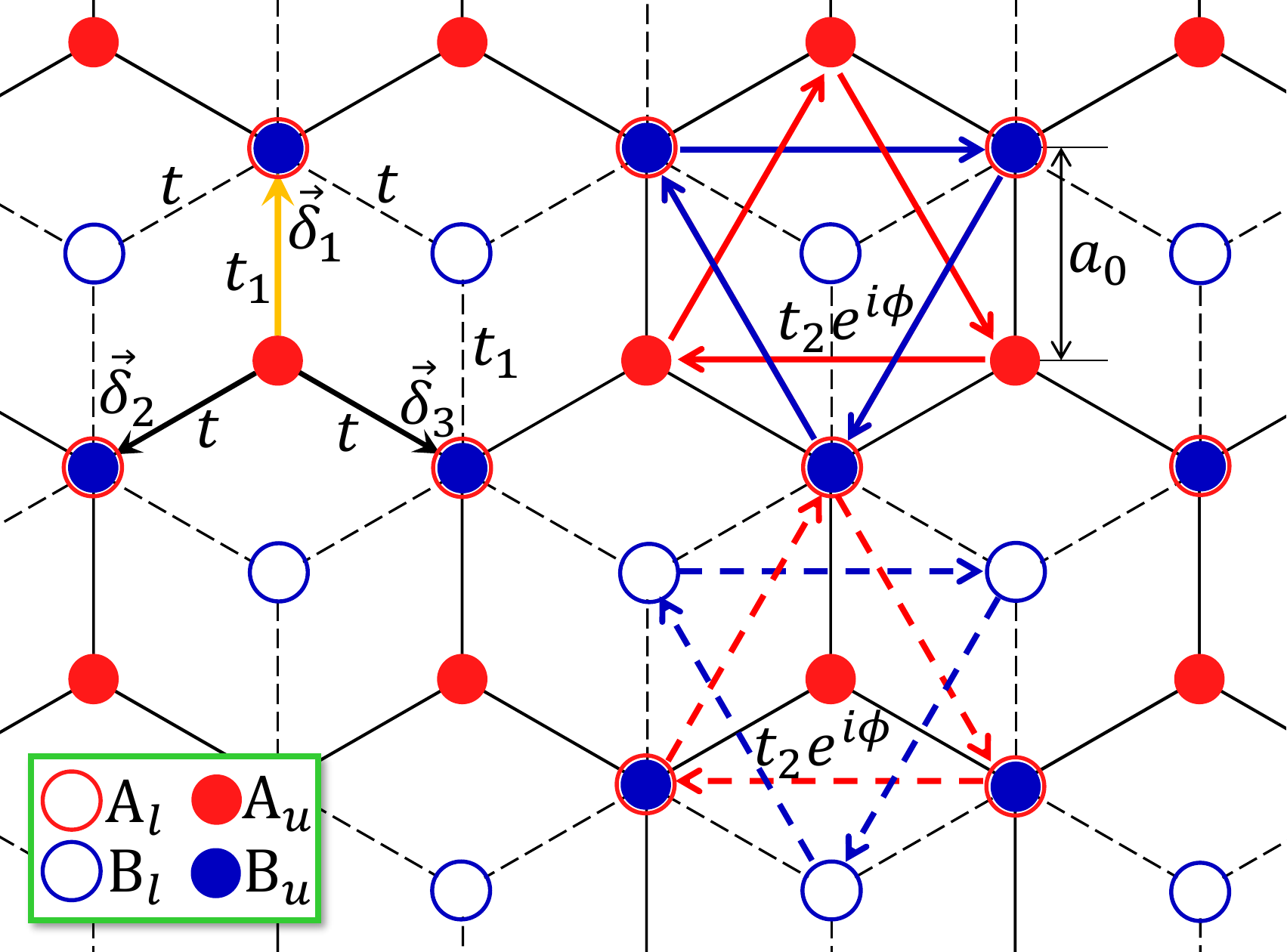}
			\subcaption{}\label{fig:lattice_hoppings}
		\end{subfigure}
		
		\caption{\raggedright A bilayer graphene is shown in (a) with the interlayer coupling $t_\perp$ between the B sublattice of upper layer and the A sublattice of lower layer. In both layers, the A and B sublattices are denoted by the red and blue filled circles. In (b), the other planar hoppings are shown. To properly see each sublattices in each layer, we have denoted the A and B sublattices in lower layer with the circles in red and blue respectively. The subscripts $l$ and $u$ in $\mathrm{A}_{l,u}$ and $\mathrm{B}_{l,u}$ refers to lower and upper layer respectively. All the bondings and NNN hoppings in the lower layers are shown by the dashed lines and dashed arrows respectively. The NN hopping strength along the $\boldsymbol{\delta}_1$ direction (shown via the yellow arrow) is $t_1$, while it is $t$ along the $\boldsymbol{\delta}_{2,3}$ directions ($\boldsymbol{\delta}_i$ are defined in text). The NNN hopping is $t_2e^{i\phi}$ ($t_2e^{-i\phi}$) for the clockwise (anti-clockwise) direction.}\label{fig:lattice_struc}
	\end{figure}
	
	In this work, we focus on a bilayer graphene with broken TRS, that is, a coupled bilayer Haldane model. The stacking of the two layers is assumed in such a way that the B subllatice of the upper layer lies exactly above the A sublattice of the lower layer. Such stacking is known as the AB stacking or the Bernal stacking. We shall see that the Chern numbers associated with various bands reveal interesting properties. For example, some of the bands possess both Chern numbers $\pm2$ and $\pm1$, while the rests are associated with Chern numbers $\pm1$. Such a scenario needs to be assessed for a band engineered system. Specifically, we wish to address the ramifications of the band deformation caused via asymmetric hopping amplitudes on the topological properties and ascertain whether such deformation induces a topological phase transition. In our bilayer model, the band engineering is incorporated via asymmetric NN hopping amplitudes in each of the layers, while the tunneling  amplitude across the layers is left unaltered.

	Our subsequent discussions have been arranged as follows. Sec. \ref{sec:model_hamiltonian} introduces the tight binding Hamiltonian of a bilayer graphene. Sec. \ref{sec:bandstructure} discusses the band structure of the system with the interlayer coupling ($t_\perp$) and the anisotropic NN hopping amplitudes ($t_1$) as parameters. Sec. \ref{sec:phase_diagram} deals with the phase diagrams that are obtained by computing the Chern numbers associated with the bands. In Sec. \ref{sec:edge_states}, the presence (or absence) of the chiral edge modes in a ribbon geometry are presented. Next, the numerical computations of the anomalous Hall conductivity are shown in Sec. \ref{sec:hall_conductivity}. Finally, a brief summary of the results are included in the concluding section (Sec. \ref{sec:conclusion}).

	\section{The Hamiltonian}\label{sec:model_hamiltonian}
	
		\begin{figure}[h]
		\captionsetup[subfigure]{labelformat=nocaption}
		\centering
		\includegraphics[width=\linewidth]{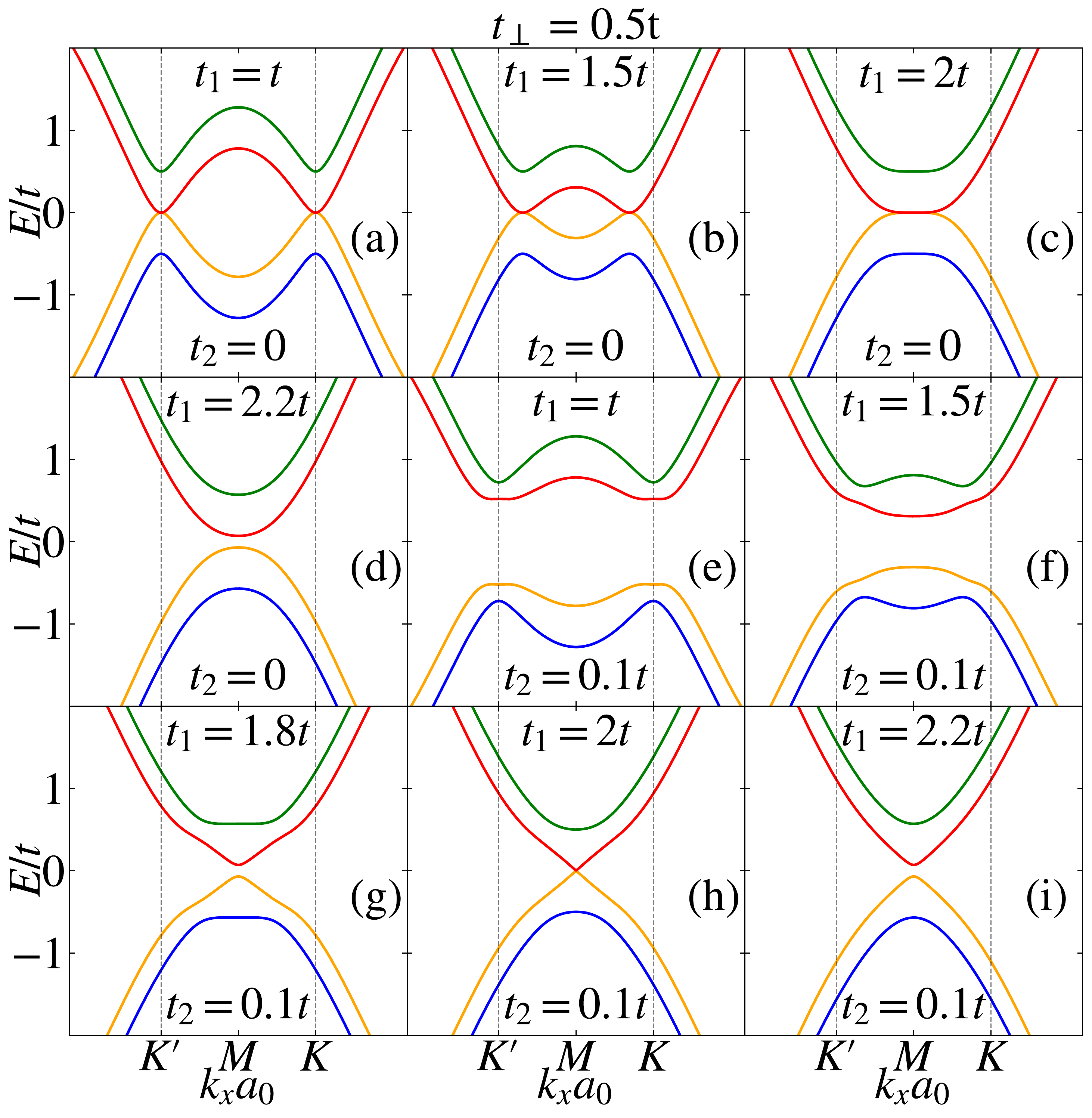}
		
		\begin{subfigure}[b]{0\textwidth}
			\subcaption{}\label{fig:band_t1_10_tc_5_t2_0}
		\end{subfigure}
		\begin{subfigure}[b]{0\textwidth}
			\subcaption{}\label{fig:band_t1_15_tc_5_t2_0}
		\end{subfigure}
		\begin{subfigure}[b]{0\textwidth}
			\subcaption{}\label{fig:band_t1_20_tc_5_t2_0}
		\end{subfigure}
		\begin{subfigure}[b]{0\textwidth}
			\subcaption{}\label{fig:band_t1_22_tc_5_t2_0}
		\end{subfigure}
		\begin{subfigure}[b]{0\textwidth}
			\subcaption{}\label{fig:band_t1_10_tc_5_t2_1}
		\end{subfigure}
		\begin{subfigure}[b]{0\textwidth}
			\subcaption{}\label{fig:band_t1_15_tc_5_t2_1}
		\end{subfigure}
		\begin{subfigure}[b]{0.\textwidth}
			\subcaption{}\label{fig:band_t1_18_tc_5_t2_1}
		\end{subfigure}
		\begin{subfigure}[b]{0.\textwidth}
			\subcaption{}\label{fig:band_t1_20_tc_5_t2_1}
		\end{subfigure}
		\begin{subfigure}[b]{0.\textwidth}
			\subcaption{}\label{fig:band_t1_22_tc_5_t2_1}
		\end{subfigure}
		\caption{\raggedright The band structure in absence of $t_2$ $(t_2=0)$ is shown along the $k_x$-axis  (at $k_ya_0 = 2\pi/3$) for (a) $t_1 = t$, (b) $t_1 = 1.5t$, (c) $t_1 = 2t$, and (d) $t_1 = 2.2t$. Similarly, the dispersions in presence of $t_2$ $(t_2=0.1t)$ are depicted for (e) $t_1 = t$, (f) $t_1 = 1.5t$, (g) $t_1 = 1.8t$, (h) $t_1 = 2t$, and (i) $t_1 = 2.2t$. The values of the other parameters are $t_\perp=0.5t$ and $\phi_l=\phi_u=\pi/2$.}
		\label{fig:band1}
	\end{figure}

	\begin{figure}[h]
		\captionsetup[subfigure]{labelformat=nocaption}
		\centering
		\includegraphics[width=\linewidth]{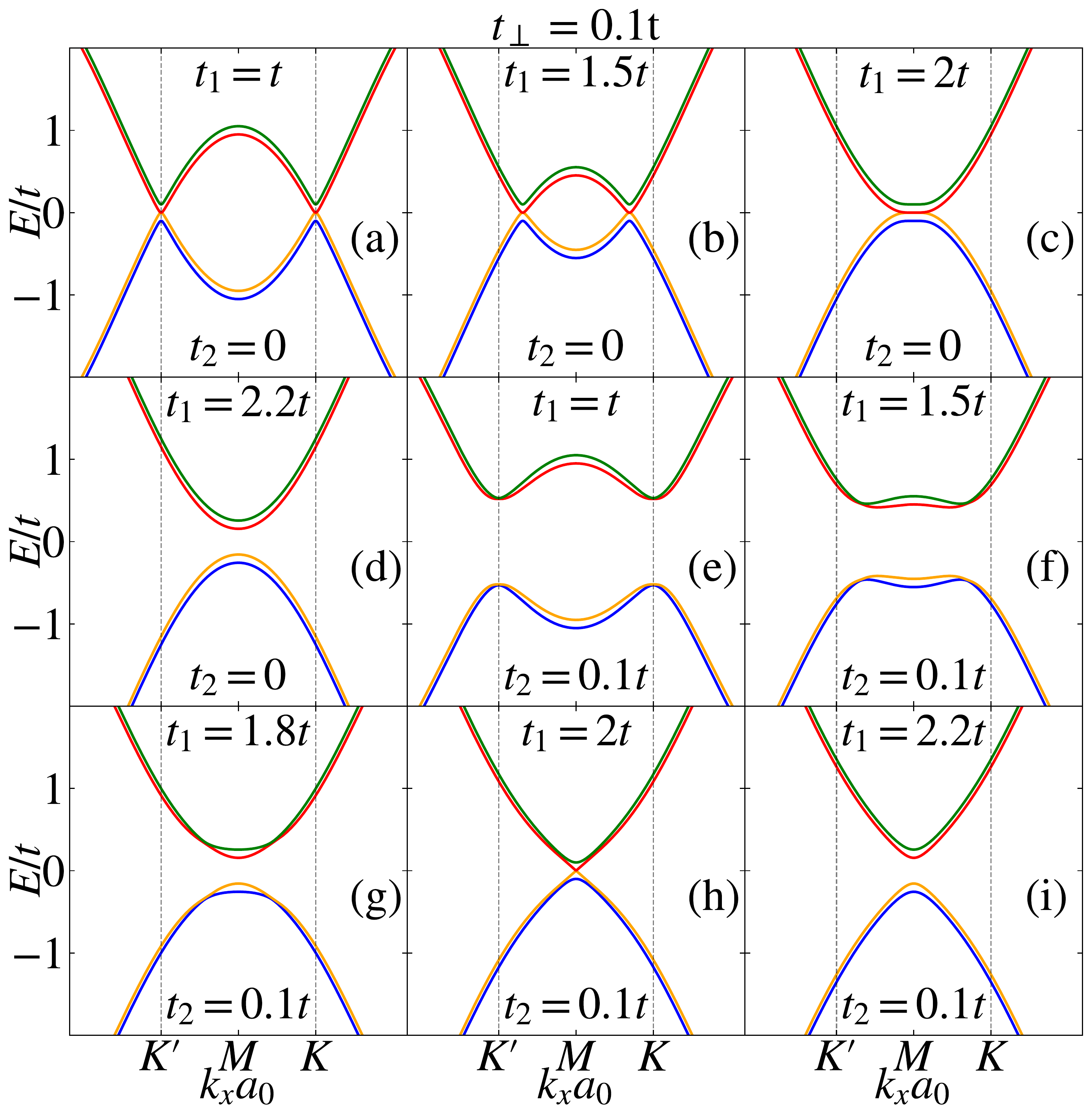}
		
		\begin{subfigure}[b]{0\textwidth}
			\subcaption{}\label{fig:band_t1_10_tc_1_t2_0}
		\end{subfigure}
		\begin{subfigure}[b]{0\textwidth}
			\subcaption{}\label{fig:band_t1_15_tc_1_t2_0}
		\end{subfigure}
		\begin{subfigure}[b]{0\textwidth}
			\subcaption{}\label{fig:band_t1_20_tc_1_t2_0}
		\end{subfigure}
		\begin{subfigure}[b]{0\textwidth}
			\subcaption{}\label{fig:band_t1_22_tc_1_t2_0}
		\end{subfigure}
		\begin{subfigure}[b]{0\textwidth}
			\subcaption{}\label{fig:band_t1_10_tc_1_t2_1}
		\end{subfigure}
		\begin{subfigure}[b]{0\textwidth}
			\subcaption{}\label{fig:band_t1_15_tc_1_t2_1}
		\end{subfigure}
		\begin{subfigure}[b]{0.\textwidth}
			\subcaption{}\label{fig:band_t1_18_tc_1_t2_1}
		\end{subfigure}
		\begin{subfigure}[b]{0.\textwidth}
			\subcaption{}\label{fig:band_t1_20_tc_1_t2_1}
		\end{subfigure}
		\begin{subfigure}[b]{0.\textwidth}
			\subcaption{}\label{fig:band_t1_22_tc_1_t2_1}
		\end{subfigure}
		\caption{\raggedright The band structure for $t_2=0$ is shown along the $k_x$-axis  (at $k_ya_0 = 2\pi/3$) for (a) $t_1 = t$, (b) $t_1 = 1.5t$, (c) $t_1 = 2t$, and (d) $t_1 = 2.2t$. While, the spectra for a non-zero $t_2$ $(t_2=0.1t)$ are depicted for
		(e) $t_1 = t$, (f) $t_1 = 1.5t$, (g) $t_1 = 1.8t$, (h) $t_1 = 2t$, and (i) $t_1 = 2.2t$. The values of $t_\perp$, $\phi_l$ and $\phi_u$ are fixed at $0.1t$, $\pi/2$ and $\pi/2$ respectively.}
		\label{fig:band2}
	\end{figure}

	A tight-binding Hamiltonian of a bilayer honeycomb lattice can be written as follows,
	\begin{align}\label{eq:ham1}
		H = & \sum_{p\in l, u}\left[ \sum_{\langle ij \rangle} t_{ij} c_i^{p\dagger} c^p_j + t_2\sum_{\langle\langle i m \rangle\rangle} e^{i\phi^{im}_p} c_i^{p\dagger} c^p_m  + \mathrm{h.c.} \right] \nonumber\\ 
		& + \left[ t_\perp\sum_{{\left\langle q, r \right\rangle}_\perp} c_q^{l\dagger} c^u_r + \mathrm{h.c.} \right]
	\end{align}
	where $c_i^{p\dagger} (c_i^p)$ is the creation (annihilation) operator corresponding to site $i$ which belongs to the layer $p$. Here $p=l, u$ represent the lower and the upper layers respectively. The first term in the right hand side denotes the nearest neighbour (NN) hopping with the amplitude $t_{ij}$ being either $t_1$ when $i$ and $j$ sites lie along the $\boldsymbol{\delta}_1 = a_0 (0, 1)$ direction, or $t$ when they lie along the $\boldsymbol{\delta}_2= a_0(\sqrt{3}/2, -1/2)$ and $\boldsymbol{\delta}_3= a_0(\sqrt{3}/2, -1/2)$ directions as shown in Fig. \ref{fig:lattice_struc}. The second term represents the complex next nearest neighbour (NNN) hopping with the amplitude $t_2$ and a phase $\phi^{im}_{l,p}$. We have labeled the Haldane flux corresponding to the lower and upper layers as $\phi^{im}_l$ and $\phi^{im}_u$ respectively. If an electron hops in the counter-clockwise direction, $\phi^{im}_{l,p}$ assumes a positive sign, while for the clockwise direction, it acquires negative sign. The third term is the hopping between the two layers with the coupling strength $t_\perp$. It should be kept in mind that the interlayer hopping is between the B sublattice on layer $u$ ($r\in \mathrm{B}_u$) and the A sublattice on layer $l$ ($q\in \mathrm{A}_l$) (AB or Bernal stacking). In our calculations, we have varied $t_1$ in both the layers from a value $t$ to $2t$ (semi-Dirac) and even considered $t_1>2t$.

	Now, we Fourier transform the Hamiltonian and write them in the four sublattice basis, namely, $\{ \mathrm{A}_l, \mathrm{B}_l, \mathrm{A}_u, \mathrm{B}_u \}$ in the following way,
	\begin{equation}\label{eq:ham_kspace}
		H(\mathbf{k}) = 
		\begin{pmatrix}
			h^+_z(\mathbf{k}, \phi_l) & h_{xy}(\mathbf{k}, t_1) & 0 & t_\perp \\
			h_{xy}^*(\mathbf{k},  t_1) & h_z^-(\mathbf{k}, \phi_l) & 0 & 0 \\
			0 & 0 & h^+_z(\mathbf{k}, \phi_u) & h_{xy}(\mathbf{k}, t_1)\\
			t_\perp & 0 & h_{xy}^*(\mathbf{k}, t_1) & h^-_z(\mathbf{k}, \phi_u) \\
		\end{pmatrix}
	\end{equation}
	 where $h_z^\pm$ are defined as, $h^+_z(\mathbf{k}, \phi_p) = h_0(\mathbf{k}, \phi_p) \pm h_z(\mathbf{k}, \phi_p$). The element $h_{xy}(\mathbf{k}, t_1)$ has the following form, $h_{xy}(\mathbf{k}, t_1) = h_x(\mathbf{k}, t_1) - i h_y(\mathbf{k}, t_1)$. The expressions for the  $h_i$s can be written as,
	 \begin{align}
	 	h_0(\mathbf{k}, \phi_p)= 2t_2\cos \phi_p \left\{2\cos\frac{\sqrt{3}k_x}{2} \cos\frac{3k_y}{2} + \cos\sqrt{3}k_x \right\}
	 \end{align}
	 \begin{align}
	 	h_z(\mathbf{k}, \phi_p)= -2t_2\sin \phi_p\left\{2\sin\frac{\sqrt{3}k_x}{2} \cos\frac{3k_y}{2} - \sin\sqrt{3}k_x \right\}
	 \end{align}
	\begin{equation}
		h_x(\mathbf{k}, t_1) = \left\{t_1 \cos k_y + 2t\cos \frac{k_y}{2} \cos\frac{\sqrt{3}k_x}{2}\right\},
	\end{equation}
	and
	\begin{equation}
		h_y(\mathbf{k}, \tilde{t}) = \left\{-t_1 \sin k_y + 2t\sin \frac{k_y}{2} \cos\frac{\sqrt{3}k_x}{2}\right\},
	\end{equation}
	
	Throughout our work, the amplitude of the NNN hopping $t_2$ is kept fixed at $0.1t$, and two different values of the interlayer hopping strength are chosen, namely, $t_\perp=0.5t$ and $t_\perp=0.1t$ \cite{phiphi_pd}. The values of $\phi_l$ and $\phi_u$ are taken such that $\phi_l=\phi_u=\pi/2$. Now, for $\phi_u=\phi_l$, we obtain the following dispersion relation,
	
	\begin{align}\label{eq:E_k_conduction}
		E^c_\pm= \left[ h_0 + \sqrt{\frac{t_\perp^2}{2} + |h_{xy}|^2 + h_z^2 \pm \frac{t_\perp}{2}\sqrt{t_\perp^2+4h_{xy}^2}} \right]
	\end{align}
	\begin{align}\label{eq:E_k_valence}
		E^v_\pm= \left[ h_0 - \sqrt{\frac{t_\perp^2}{2} + |h_{xy}|^2 + h_z^2 \pm \frac{t_\perp}{2}\sqrt{t_\perp^2+4h_{xy}^2}} \right]
	\end{align}
	where $E^c_\pm$ denote the two conduction bands and $E^v_\pm$ are the two valence bands for a bilayer.
	
	\begin{figure}[h]
	\captionsetup[subfigure]{labelformat=nocaption}
	\centering
	\includegraphics[width=\linewidth]{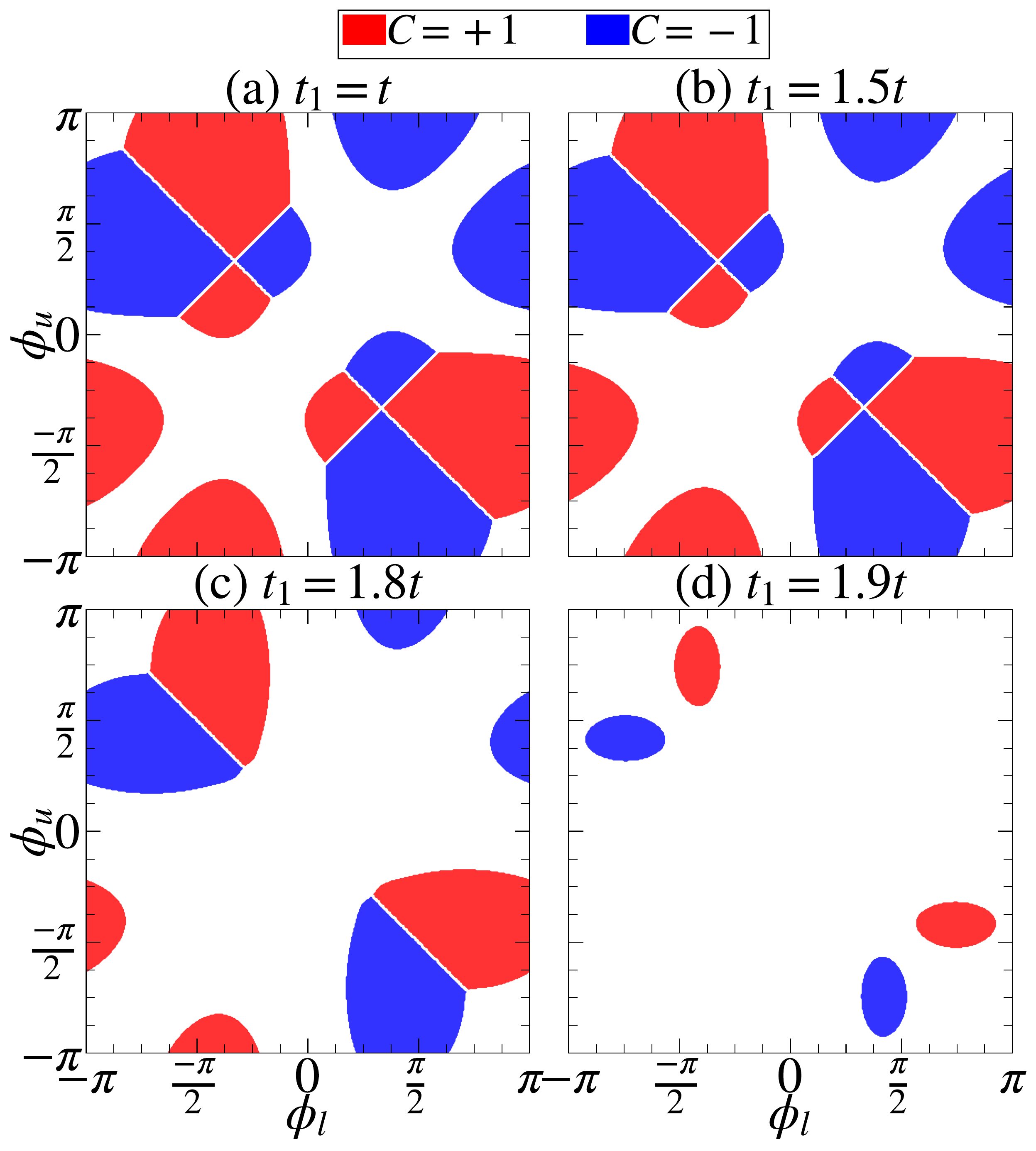}
	
	\begin{subfigure}[b]{0\textwidth}
		\subcaption{}\label{fig:pd_phi_t1_1_tcp_0.5_bd_1}
	\end{subfigure}
	\begin{subfigure}[b]{0\textwidth}
		\subcaption{}\label{fig:pd_phi_t1_1.5_tcp_0.5_bd_1}
	\end{subfigure}
	\begin{subfigure}[b]{0\textwidth}
		\subcaption{}\label{fig:pd_phi_t1_1.8_tcp_0.5_bd_1}
	\end{subfigure}
	\begin{subfigure}[b]{0\textwidth}
		\subcaption{}\label{fig:pd_phi_t1_1.9_tcp_0.5_bd_1}
	\end{subfigure}
	
	\caption{\raggedright The phase diagrams corresponding to the lowest occupied band, that is, band-v1 is presented for $t_\perp = 0.5t$. The white regions denote the trivial phase with Chern number as zero, while the colored regions indicate the non-trivial phase with the non-zero Chern numbers. The non-zero values are indicated at the top of the figure.}
	\label{fig:pd_phiphi_tcp_0.5_bd_1}
\end{figure}

\begin{figure}[h]
	\captionsetup[subfigure]{labelformat=nocaption}
	\centering
	\includegraphics[width=\linewidth]{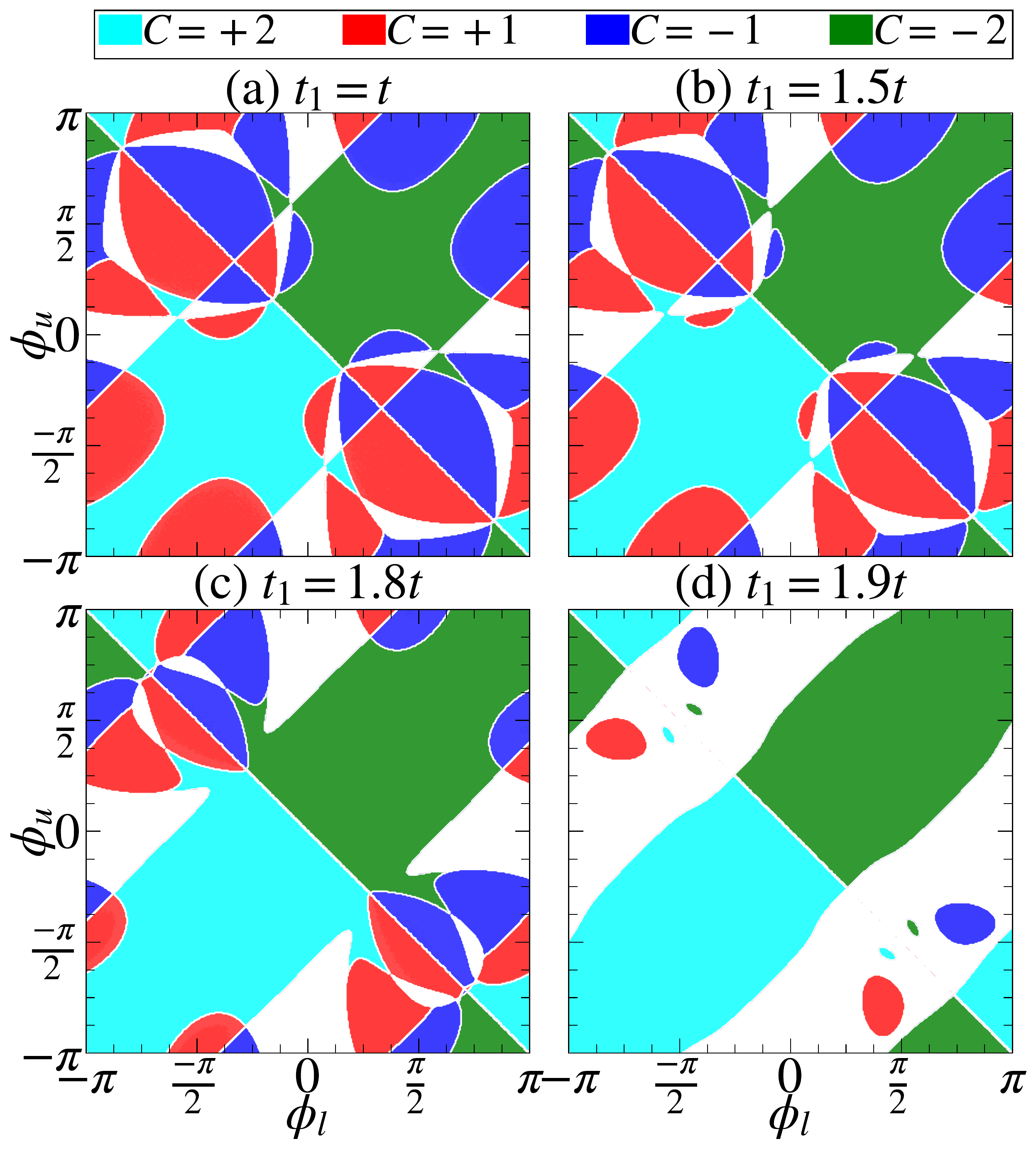}
	
	\begin{subfigure}[b]{0\textwidth}
		\subcaption{}\label{fig:pd_phi_t1_1_tcp_0.5_bd_2}
	\end{subfigure}
	\begin{subfigure}[b]{0\textwidth}
		\subcaption{}\label{fig:pd_phi_t1_1.5_tcp_0.5_bd_2}
	\end{subfigure}
	\begin{subfigure}[b]{0\textwidth}
		\subcaption{}\label{fig:pd_phi_t1_1.8_tcp_0.5_bd_2}
	\end{subfigure}
	\begin{subfigure}[b]{0\textwidth}
		\subcaption{}\label{fig:pd_phi_t1_1.9_tcp_0.5_bd_2}
	\end{subfigure}
	
	\caption{\raggedright The phase diagrams corresponding to band-v2 is presented for $t_\perp = 0.5t$. The white regions denote the trivial phases with zero Chern number, while the colored regions indicate the non-trivial phase with the non-zero Chern numbers. Again the values are indicated at the top of the figure.}
	\label{fig:pd_phiphi_tcp_0.5_bd_2}
\end{figure}	

\section{Spectral properties}\label{sec:bandstructure}

	In this section, we discuss how the spectral properties evolve as we interpolate between the Dirac and the semi-Dirac limits. We show the band structure for two different values of $t_\perp$. The first one is for $t_\perp = 0.5t$ as shown in Fig. \ref{fig:band1}. As can be seen, there are four bands which we have labeled as follows. The upper conduction band is labeled as band-c1, while the lower conduction band is band-c2. Similarly, the lower and the upper valence bands are labeled as band-v1 and band-v2 respectively. When $t_2 = 0$ (no Haldane flux), band-c2 and band-v2 touch each other at the Fermi level at the $\mathbf{K}$ and $\mathbf{K^\prime}$ points (see Figs. \ref{fig:band_t1_10_tc_5_t2_0}-\ref{fig:band_t1_22_tc_5_t2_0}). These points are referred to as the Dirac points. Further, with the increase in the value of $t_1$, we deviate from the Dirac limit, and the band touching points move close to each other which finally merge at $t_1 = 2t$. Beyond this value, that is, for $t_1>2t$, a gap opens up at the $\mathbf{M}$ point.
	Now, if we switch on $t_2$ (see Figs. \ref{fig:band_t1_10_tc_5_t2_1}-\ref{fig:band_t1_22_tc_5_t2_1}), the spectral gap remains open for $t\leq t_1 <2t$ and $t_1>2t$, while the gap vanishes exactly at the semi-Dirac limit, namely, $t_1 = 2t$. The gap closing scenario of the bilayer graphene is thus similar to the case of single layer graphene, where the energy gap between the conduction and valence band vanishes at the semi-Dirac limit, that is, at $t_1 = 2t$ \cite{mondal2021}.
	
	Further, we have presented band structure in Fig. \ref{fig:band2} for a smaller value of $t_\perp$, namely, $t_\perp = 0.1t$. It is obvious from Eqs. \ref{eq:E_k_conduction} and \ref{eq:E_k_valence} that the separation among the conduction bands (band-c1 and band-c2) and that among the valence bands (band-v1 and band-v2) decreases with decrease in $t_\perp$. Moreover, the low energy dispersions of band-c2 and band-v2 about the band touching points have a linear behaviour which was quadratic for $t_\perp = 0.5t$. Thus, the massive electrons become progressively massless as we lower the value of $t_\perp$. Further, with the decrease in $t_\perp$, the spectral gap between band-c2 and band-v2 increases. For example, when $t_\perp = 0.1t$ the band gap is $\Delta E_g \simeq 1.0390t$ and $0.3124t$ for $t_1= t$ and $1.8t$ respectively. While for $t_\perp = 0.5t$, $\Delta E_g \simeq 1.0335t$ and $0.1406t$ for $t_1= t$ and $1.8t$ respectively. Thus the difference in energy is more noticeable as we move towards the semi-Dirac limit, that is, at large values of $t_1$.

\section{Chern number and phase diagram}\label{sec:phase_diagram}

	\begin{figure}[h]
		\captionsetup[subfigure]{labelformat=nocaption}
		\centering
		\includegraphics[width=\linewidth]{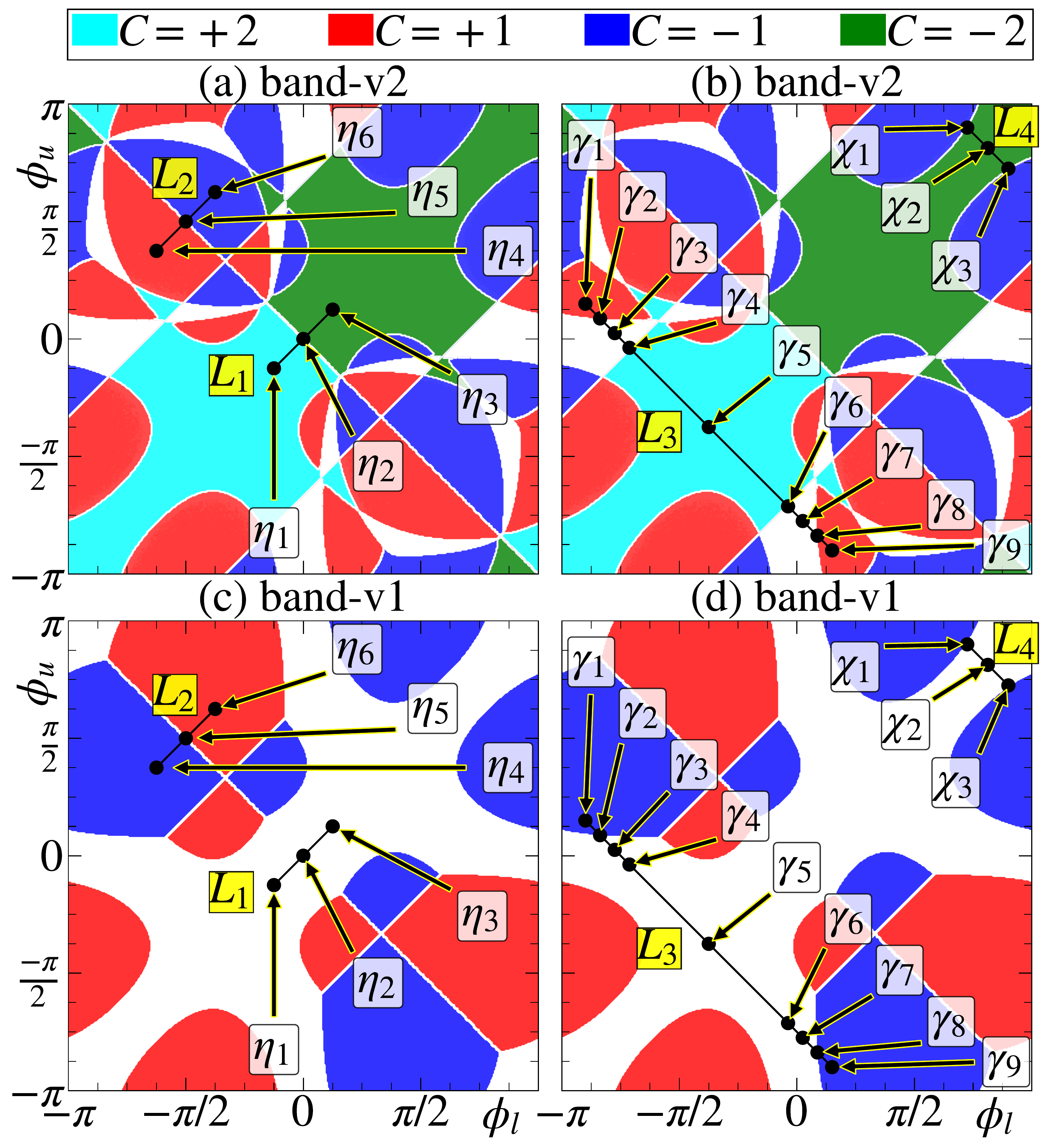}
		
		\begin{subfigure}[b]{0\textwidth}
			\subcaption{}\label{fig:pd11_subfig}
		\end{subfigure}
		\begin{subfigure}[b]{0\textwidth}
			\subcaption{}\label{fig:pd12_subfig}
		\end{subfigure}
		\begin{subfigure}[b]{0\textwidth}
			\subcaption{}\label{fig:pd21_subfig}
		\end{subfigure}
		\begin{subfigure}[b]{0\textwidth}
			\subcaption{}\label{fig:pd22_subfig}
		\end{subfigure}
		
		\caption{\raggedright The phase diagrams corresponding to band-v2 are shown in (a) and (b), and those for band-v1 are presented in (c) and (d). In (a) and (c), the $\eta$ points are used and shown along the $L_1$ and $L_2$ lines, whereas the $\gamma$ and $\chi$ points are marked along the $L_3$ and $L_4$ lines in both (b) and (d). Along those lines multiple phase transitions occur. For example, along $L_3$, the Chern number corresponding to band-v2 has values $+1$, $0$, $+2$, $0$, $+1$ at the points $\gamma_1$, $\gamma_3$, $\gamma_5$, $\gamma_7$ and $\gamma_9$ respectively. The phase transitions take place at $\gamma_2$, $\gamma_4$, $\gamma_6$ and $\gamma_8$, where band-v2 touches either band-v1 or band-c2. The values of $t_\perp$ and $t_1$ are taken as $0.5t$ and $t_1$ respectively.}
		\label{fig:pds_bd}
	\end{figure}
	
	In this section, we calculate the Chern number as a function of the Haldane flux of the two layers. Owing to the broken TRS, the bands possess non-zero Chern numbers, which can be calculated by integrating the Berry curvature over the BZ \cite{thouless, avron1988}.
	\begin{eqnarray}\label{eq:chern_number}
		C & = & \frac{1}{2\pi}\int\int_{\mathrm{BZ}}\Omega(k_x, k_y)\mathrm{d}k_x \mathrm{d}k_y
	\end{eqnarray}
	where $\Omega(k_x, k_y)$ is the $z$-component of the Berry curvature \cite{liu2016}, which is obtained from the following relation.
	\begin{eqnarray}\label{eq:berry_curv}
		\Omega(k_x, k_y) = -2i\mathrm{Im}\left[\left< \frac{\partial\psi(k_x, k_y)}{\partial k_x} \right.\left|\frac{\partial\psi(k_x, k_y)}{\partial k_y}\right>\right]
	\end{eqnarray}
	where $\psi(k_x, k_y)$ is the periodic part of the Bloch wave corresponding to the Hamiltonian defined in Eq. \ref{eq:ham_kspace}, and $\mathrm{Im}$ denotes the imaginary part. Hence, we calculate the Chern numbers as a function of the fluxes $\phi_l$ and $\phi_u$ corresponding to the lower and the upper layers respectively for various values of $t_1$ as shown in Fig. \ref{fig:pd_phiphi_tcp_0.5_bd_1}. Here, the value of $t_\perp$ is chosen to be $0.5t$ and the phase diagrams shown correspond to band-v1. We have denoted the Chern insulating regions by two colors. The regions in red denote $C = +1$ phase, while the blue ones denote $C = -1$ phases. The trivial phases with $C=0$ are shown by the white regions. It is evident from Fig. \ref{fig:pd_phi_t1_1_tcp_0.5_bd_1} that the areas of the Chern insulating regions are maximum for $t_1=t$ (Dirac case). An engineering of the band structure, that is, with the increase in the value of $t_1$, the area of the topological regions (called as the Chern lobes) gradually shrink. We have shown the phase diagram till a certain value, namely, $t_1 = 1.9t$ (see Fig. \ref{fig:pd_phi_t1_1.9_tcp_0.5_bd_1}), beyond which the topological regions can hardly be seen. When $t_1$ becomes equal to $2t$, the Chern number ($C$) vanishes completely for all values of $\phi_l$ and $\phi_u$ owing to a gapless scenario between band-c2 and band-v2. Although band-v1 remains separated from the band-v2, the Chern number still vanishes. For $t_1>2t$, a gap opens up, however, the Chern numbers continue to be zero, and thus the gap is trivial. 
	
	Further, we have presented the phase diagrams corresponding to band-v2 (the one closer to the Fermi level) in Fig. \ref{fig:pd_phiphi_tcp_0.5_bd_2}. As can be seen, additional phases with higher Chern number ($C=\pm 2$) appear. We have denoted the $C = +2$ and $C = -2$ phase with cyan and green colors respectively. The red and blue colors continue to denote $C=+1$ and $C=-1$ phases respectively. Thus, both $C = \pm2$ and $C = \pm 1$ phases occur at different parameter values in the same phase diagram. Further, the topological regions shrinks with the increase in $t_1$, and finally vanishes at $t_1 = 2t$, where the gap between the band-v2 and band-c2 vanishes. For $t_1>2t$, the gap reopens, but the Chern number remains zero for all values of $\phi_l$ and $\phi_u$. The phase diagrams for band-c1 and band-c2 are identical in shape to those of band-v1 and band-v2 respectively, except the Chern numbers have opposite signs. 
	
\begin{figure*}[h]
		\centering
		
		\begin{subfigure}[b]{0.16\textwidth}
			\includegraphics[width=\textwidth]{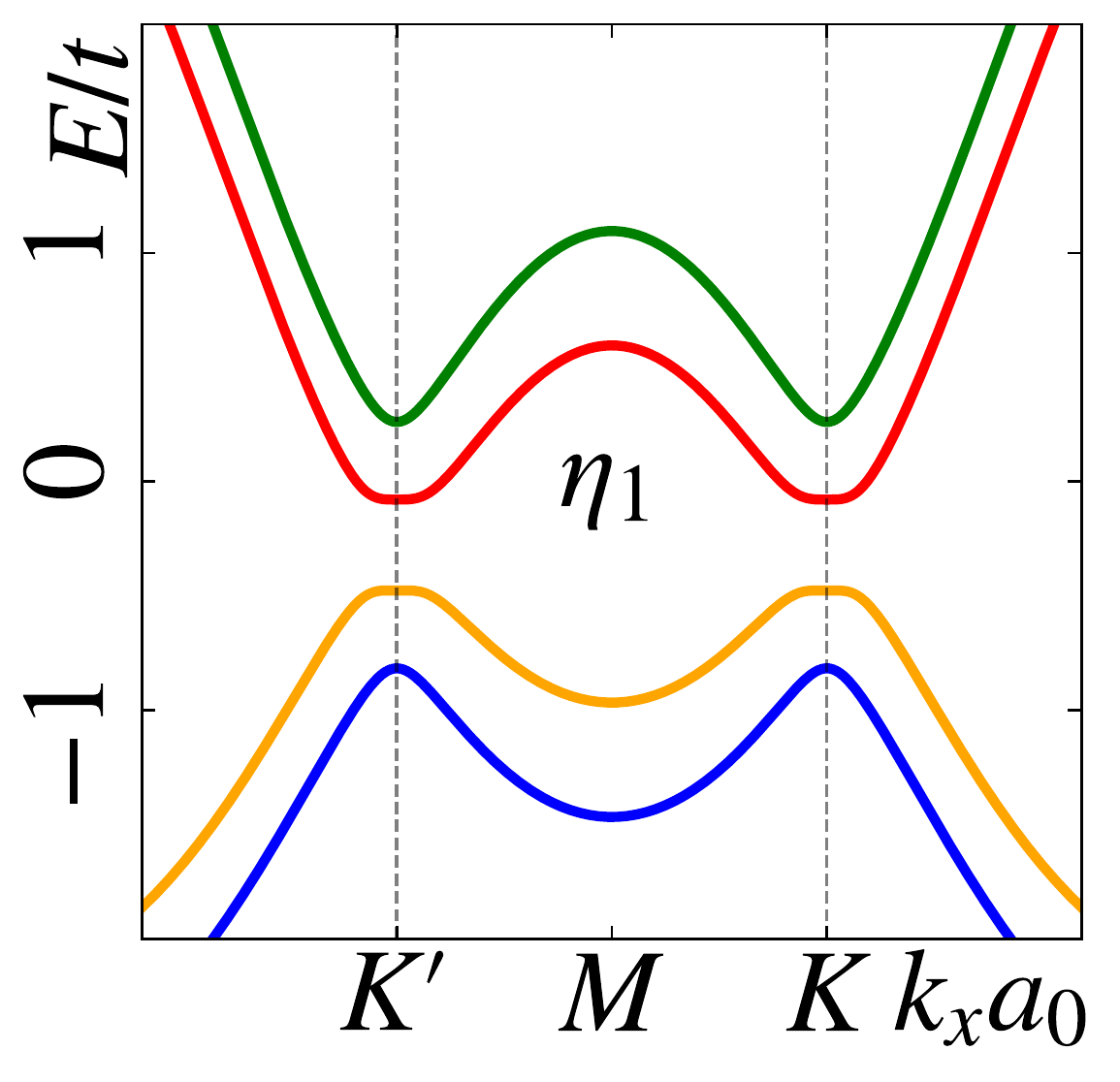}
			\subcaption{}\label{fig:band_eta1}
		\end{subfigure}
		\begin{subfigure}[b]{0.16\textwidth}
			\includegraphics[width=\textwidth]{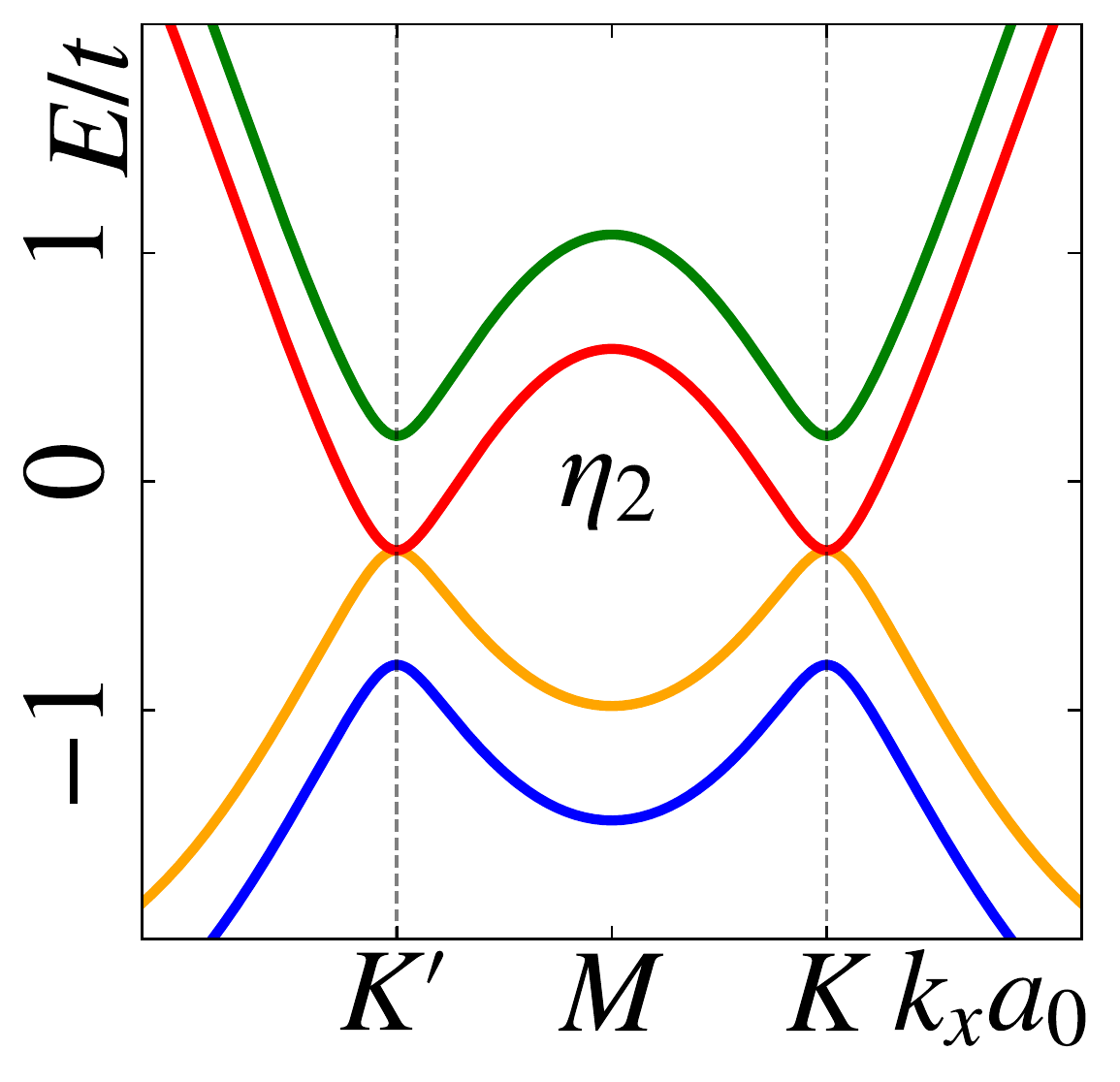}
			\subcaption{}\label{fig:band_eta2}
		\end{subfigure}
		\begin{subfigure}[b]{0.16\textwidth}
			\includegraphics[width=\textwidth]{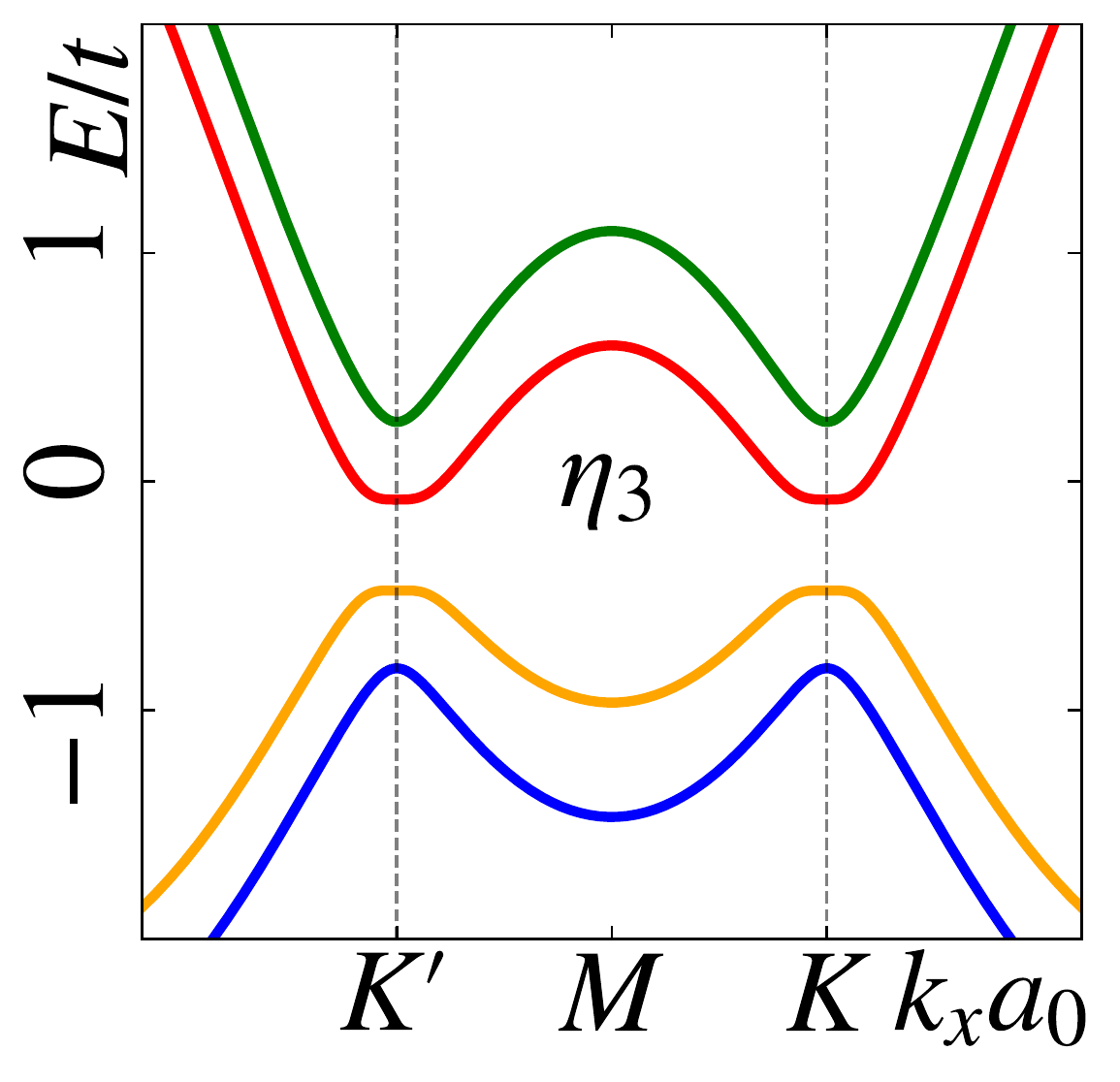}
			\subcaption{}\label{fig:band_eta3}
		\end{subfigure}
		\begin{subfigure}[b]{0.16\textwidth}
			\includegraphics[width=\textwidth]{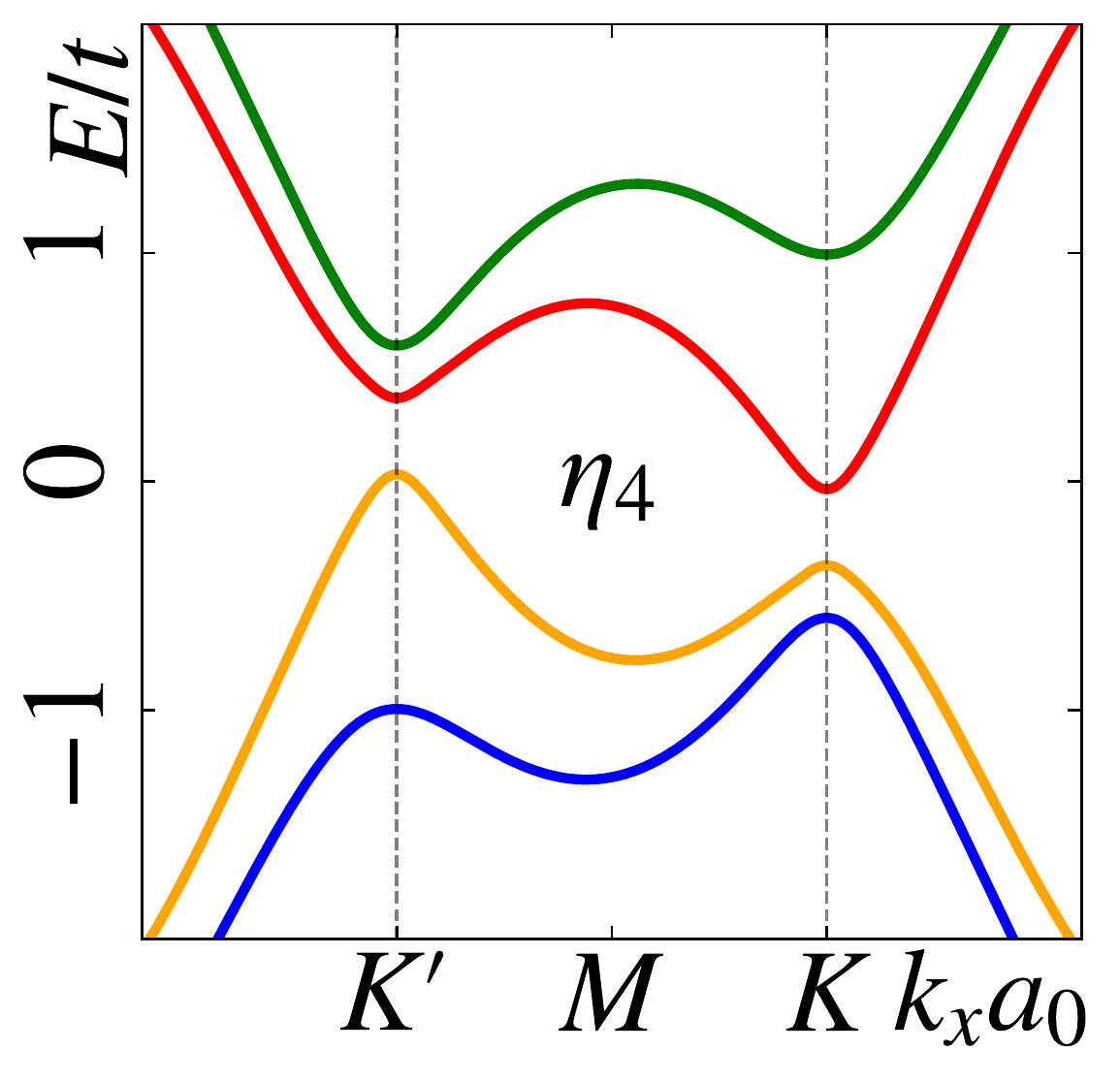}
			\subcaption{}\label{fig:band_eta4}
		\end{subfigure}
		\begin{subfigure}[b]{0.16\textwidth}
			\includegraphics[width=\textwidth]{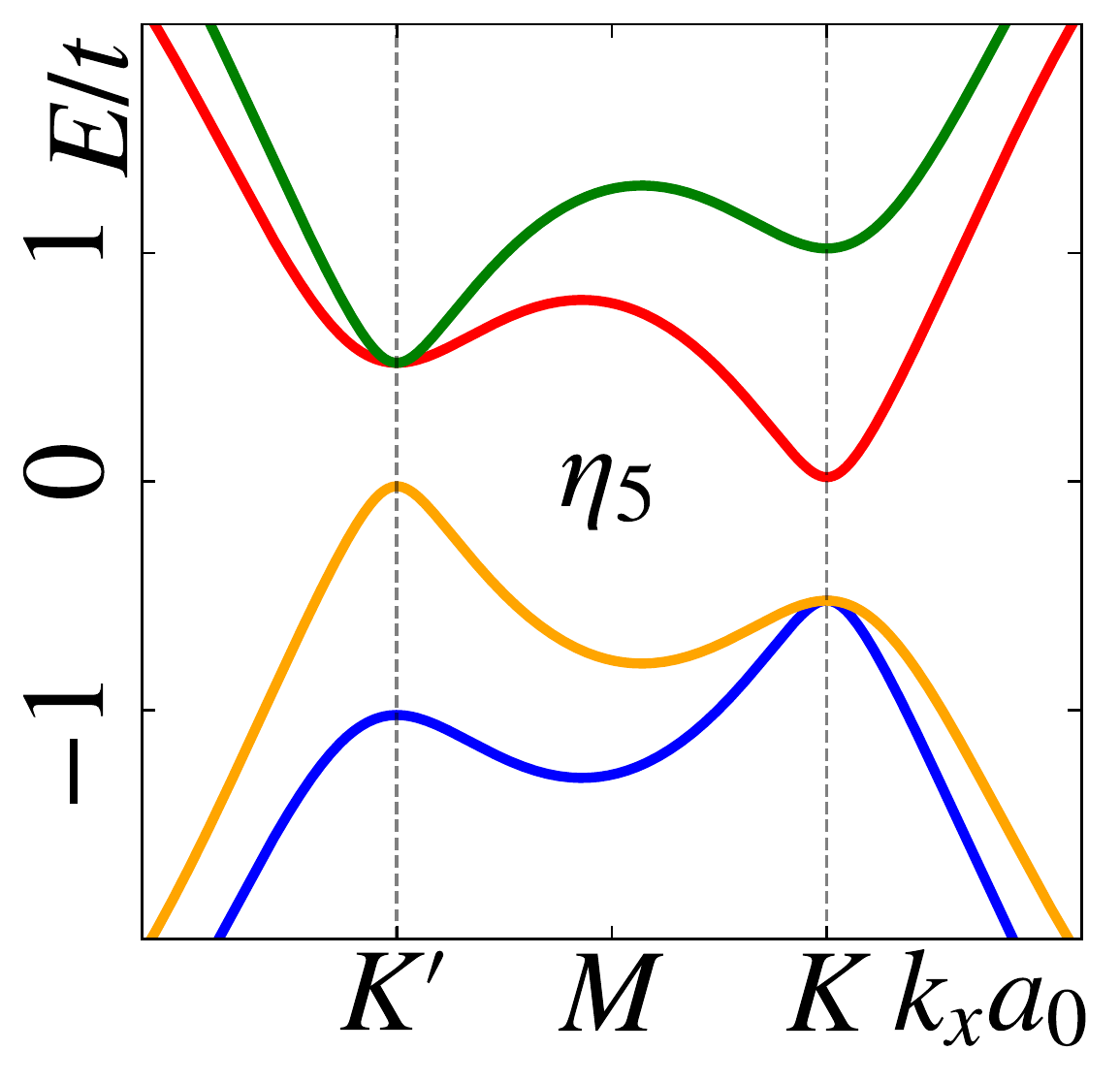}
			\subcaption{}\label{fig:band_eta5}
		\end{subfigure}
		\begin{subfigure}[b]{0.16\textwidth}
			\includegraphics[width=\textwidth]{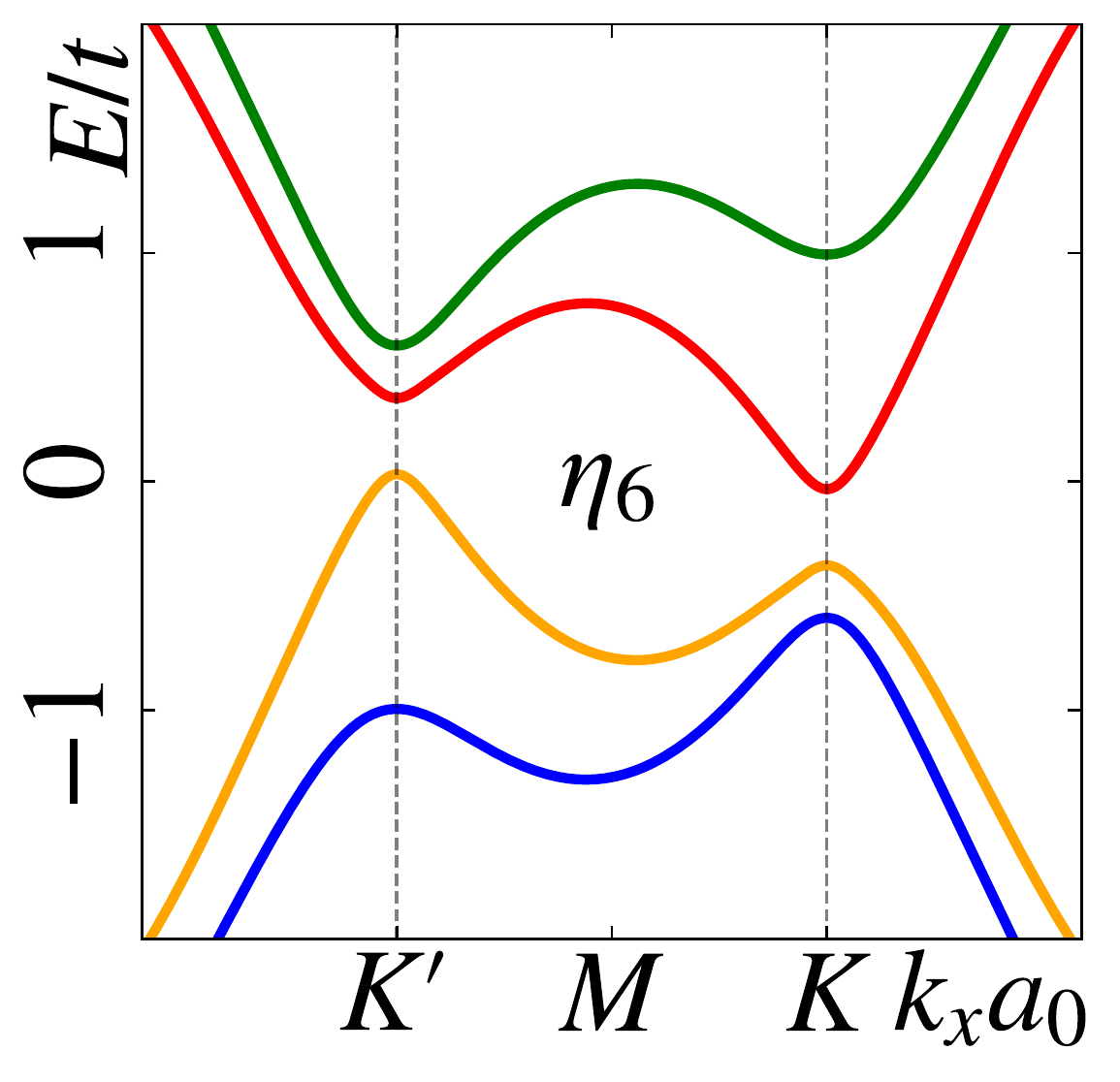}
			\subcaption{}\label{fig:band_eta6}
		\end{subfigure}
		\begin{subfigure}[b]{0.16\textwidth}
			\includegraphics[width=\textwidth]{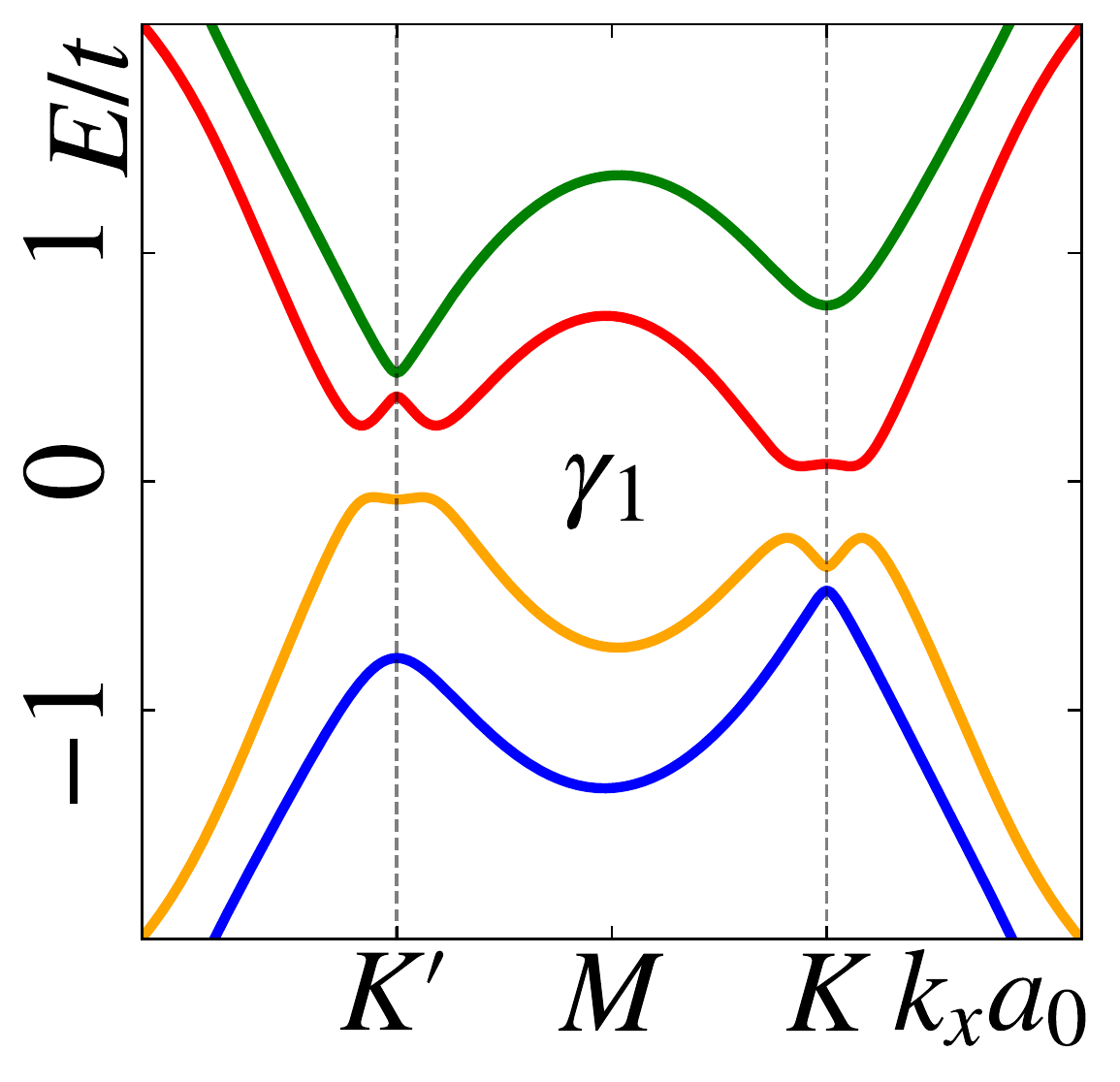}
			\subcaption{}\label{fig:band_gam1}
		\end{subfigure}
		\begin{subfigure}[b]{0.16\textwidth}
			\includegraphics[width=\textwidth]{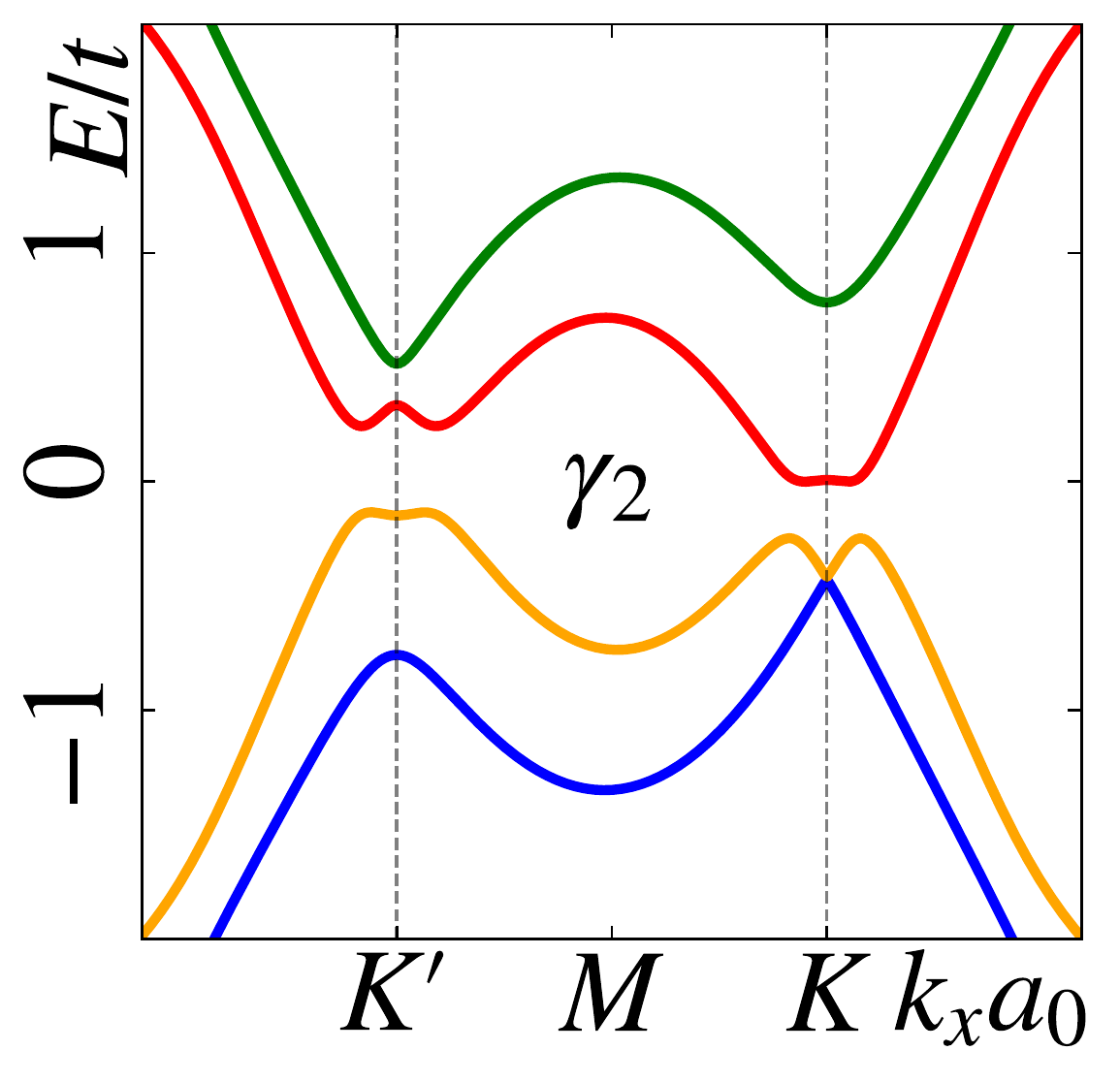}
			\subcaption{}\label{fig:band_gam2}
		\end{subfigure}
		\begin{subfigure}[b]{0.16\textwidth}
			\includegraphics[width=\textwidth]{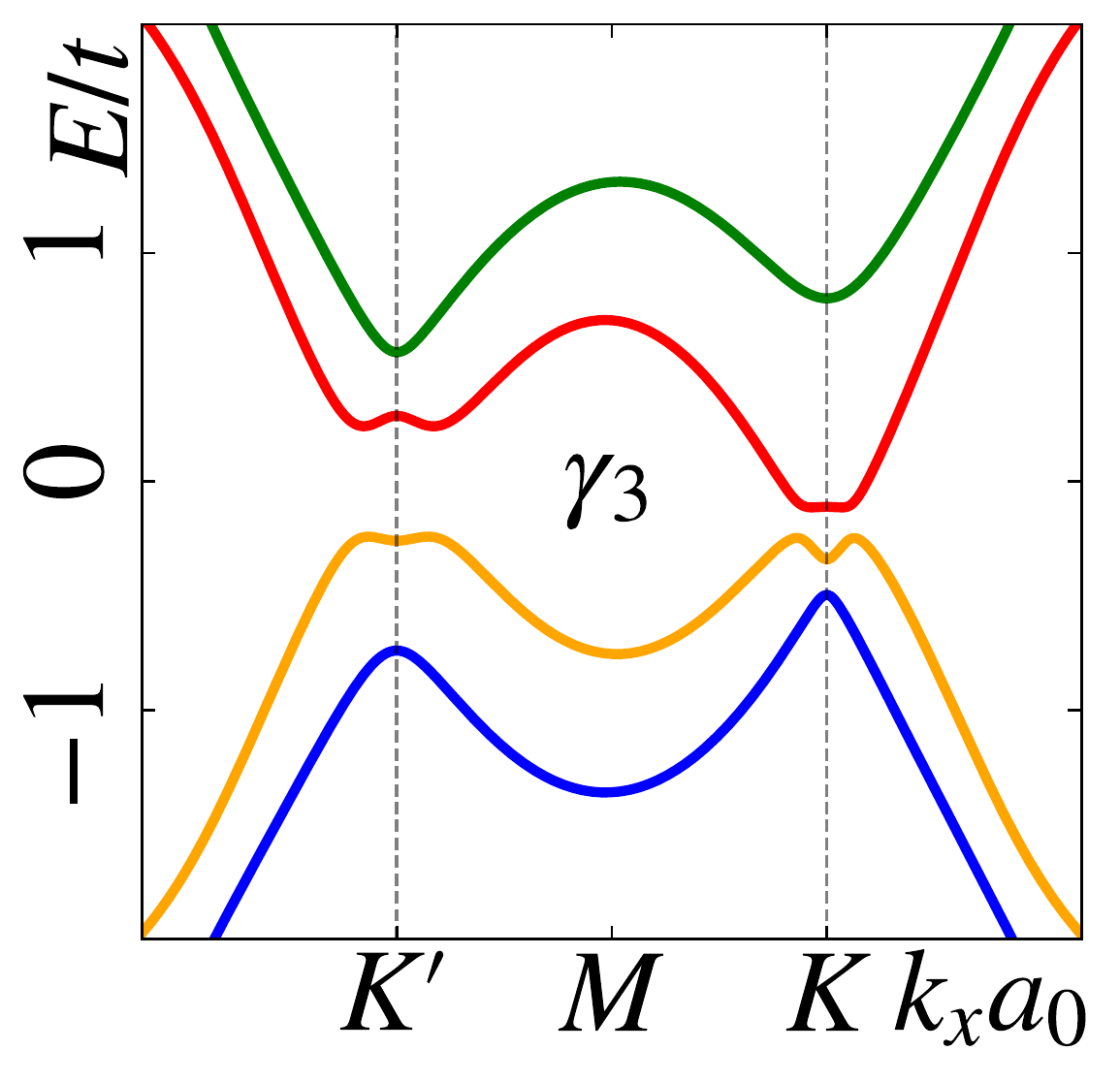}
			\subcaption{}\label{fig:band_gam3}
		\end{subfigure}
		\begin{subfigure}[b]{0.16\textwidth}
			\includegraphics[width=\textwidth]{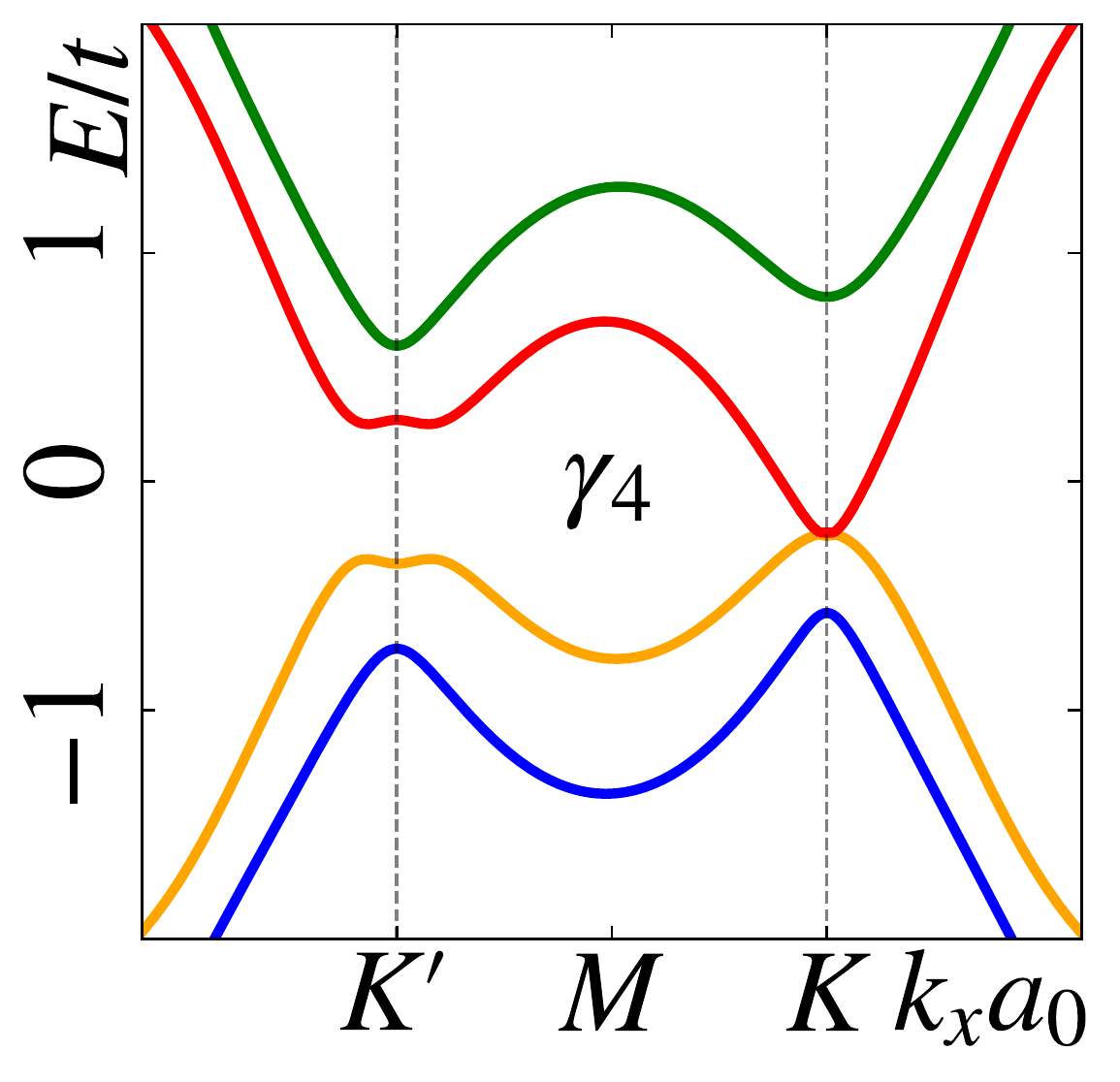}
			\subcaption{}\label{fig:band_gam4}
		\end{subfigure}
		\begin{subfigure}[b]{0.16\textwidth}
			\includegraphics[width=\textwidth]{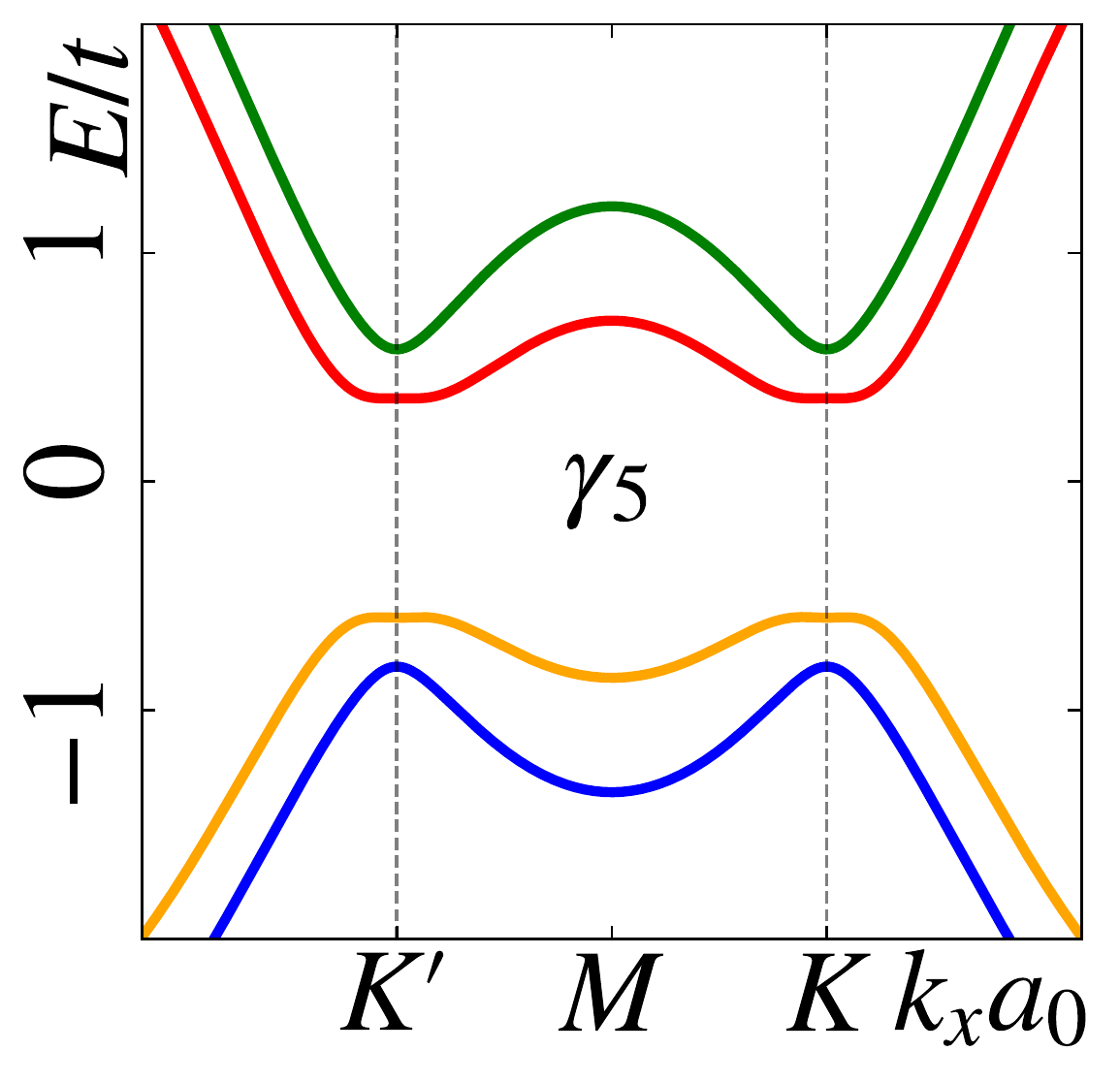}
			\subcaption{}\label{fig:band_gam5}
		\end{subfigure}
		\begin{subfigure}[b]{0.16\textwidth}
			\includegraphics[width=\textwidth]{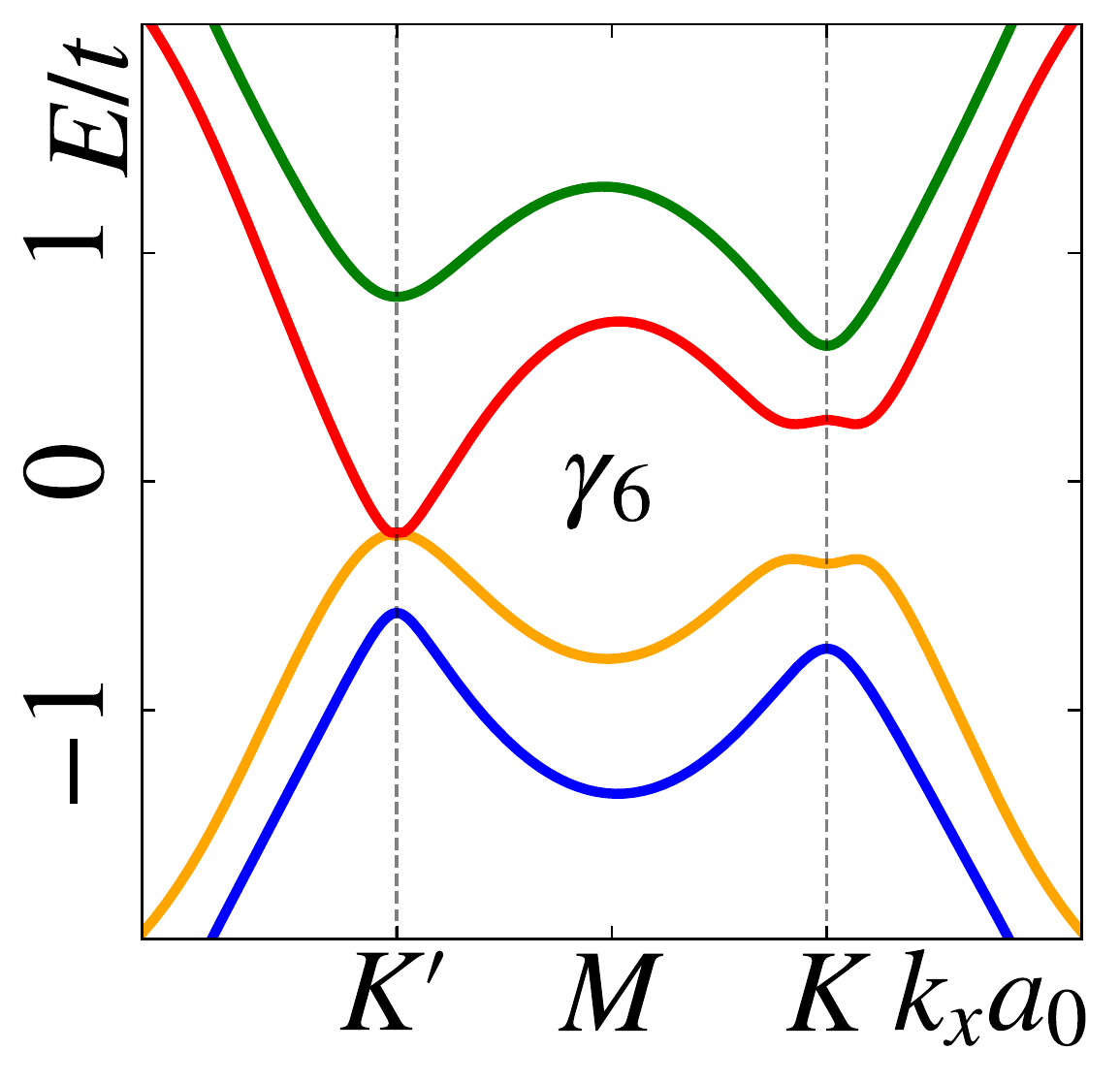}
			\subcaption{}\label{fig:band_gam6}
		\end{subfigure}
		\begin{subfigure}[b]{0.16\textwidth}
			\includegraphics[width=\textwidth]{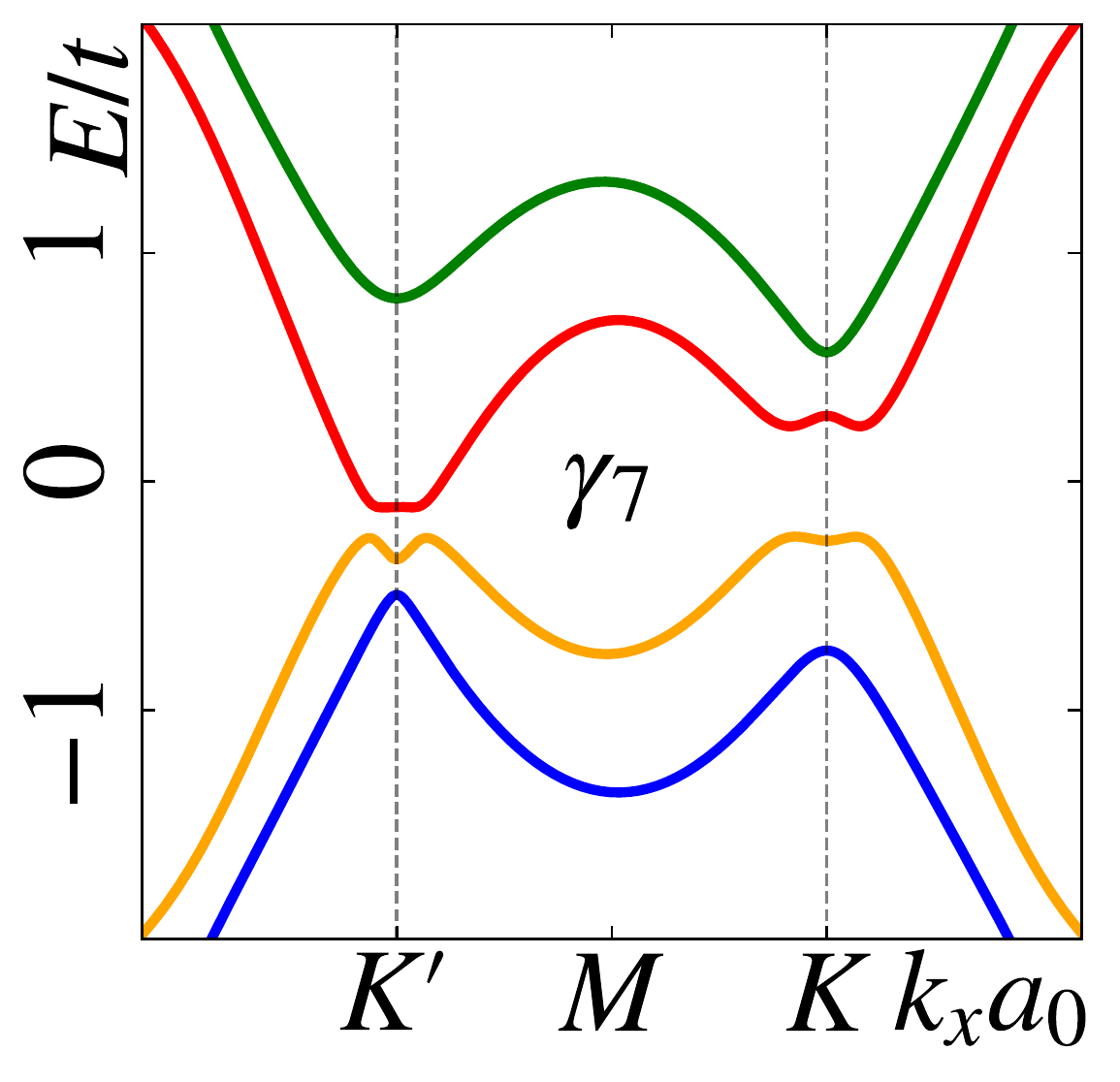}
			\subcaption{}\label{fig:band_gam7}
		\end{subfigure}
		\begin{subfigure}[b]{0.16\textwidth}
			\includegraphics[width=\textwidth]{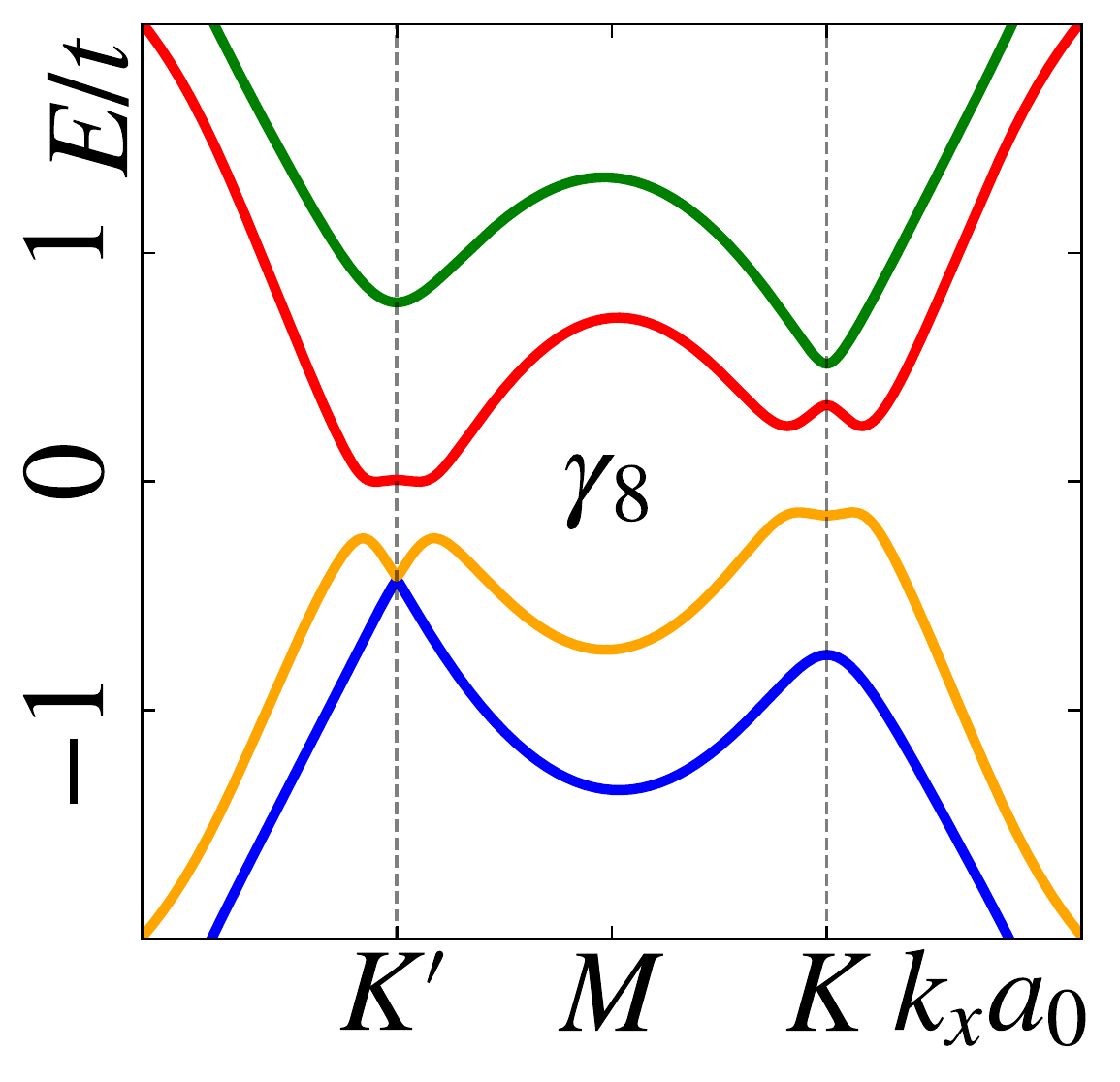}
			\subcaption{}\label{fig:band_gam8}
		\end{subfigure}
		\begin{subfigure}[b]{0.16\textwidth}
			\includegraphics[width=\textwidth]{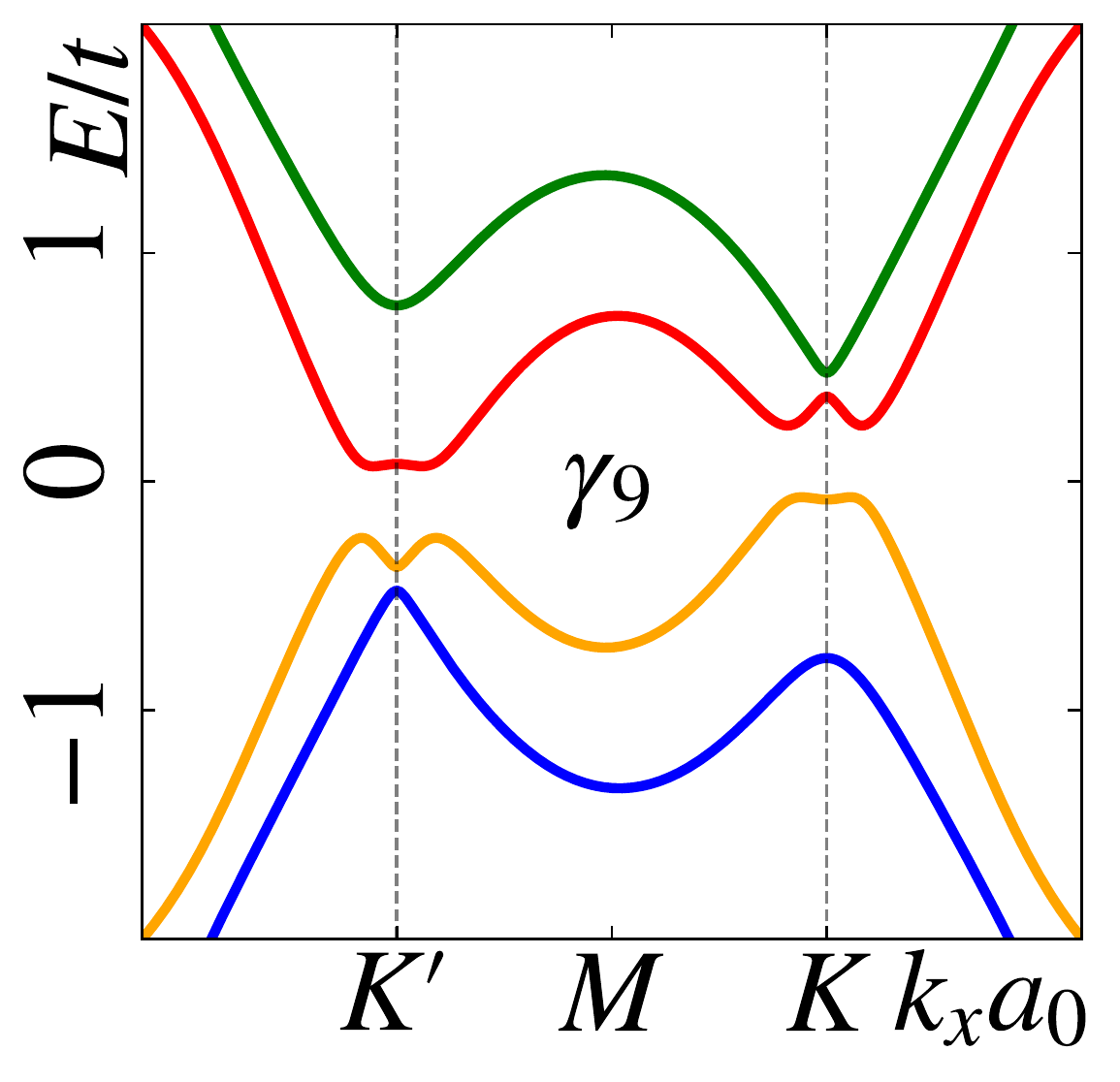}
			\subcaption{}\label{fig:band_gam9}
		\end{subfigure}
		\begin{subfigure}[b]{0.16\textwidth}
			\includegraphics[width=\textwidth]{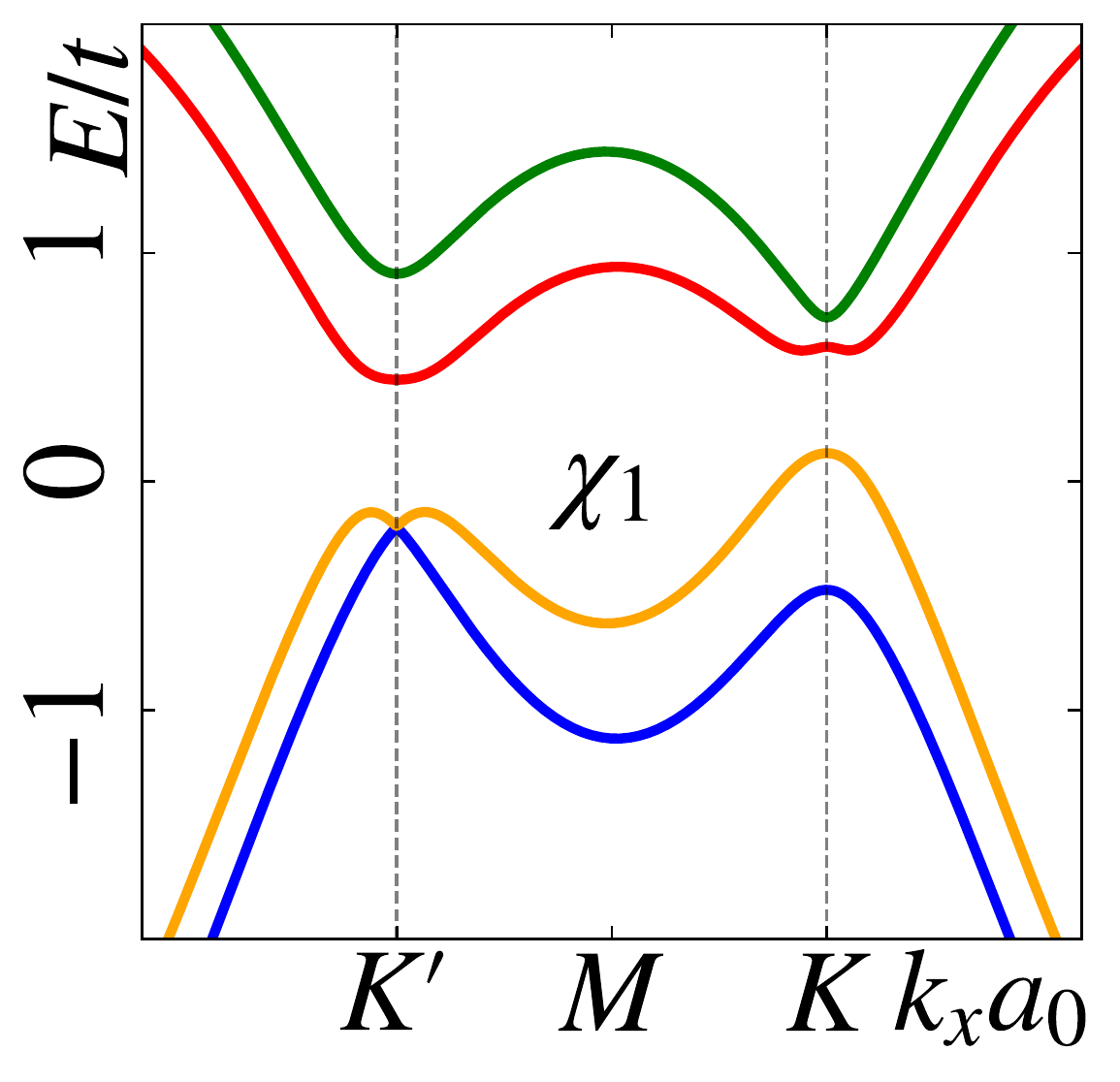}
			\subcaption{}\label{fig:band_chi1}
		\end{subfigure}
		\begin{subfigure}[b]{0.16\textwidth}
			\includegraphics[width=\textwidth]{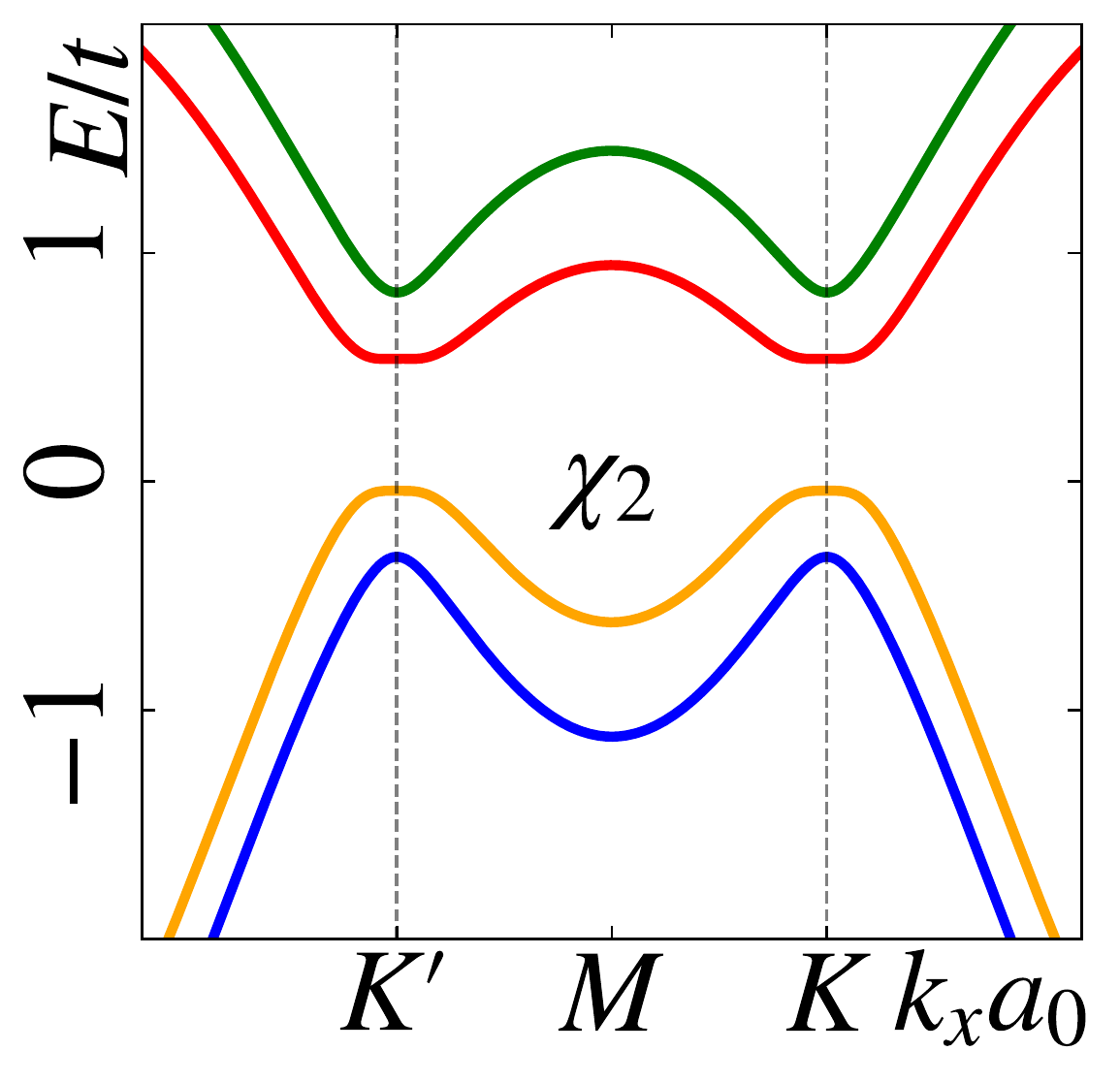}
			\subcaption{}\label{fig:band_chi2}
		\end{subfigure}
		\begin{subfigure}[b]{0.16\textwidth}
			\includegraphics[width=\textwidth]{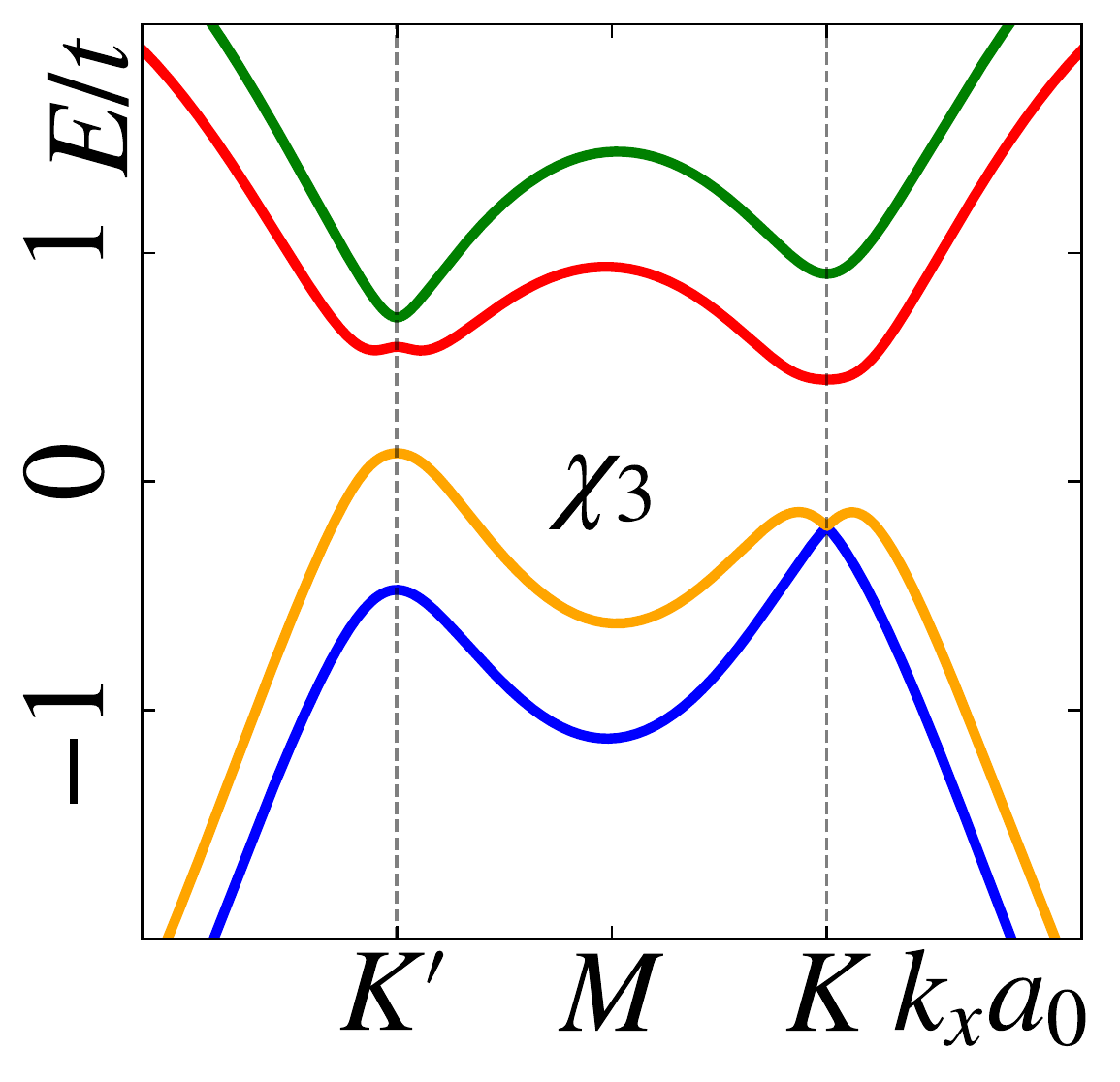}
			\subcaption{}\label{fig:band_chi3}
		\end{subfigure}
		
		\caption{\raggedright The band structures corresponding to the points $\eta_1$-$\eta_6$ (shown in Figs. \ref{fig:pd11_subfig} and \ref{fig:pd21_subfig}) are depicted in (c)-(h). The spectra for the points $\gamma_1$-$\gamma_9$ (shown in Figs. \ref{fig:pd11_subfig} and \ref{fig:pd21_subfig}) are shown in (i)-(q), and for the points $\chi_1$-$\chi_3$ are presented in (p)-(r). The values of $t_1$ and $t_\perp$ for all the band structures are kept fixed at $t$ and $0.5t$ respectively.}
		\label{fig:pds_and_bands}
	\end{figure*}

	In order to visualize the gap closing scenario corresponding to different phase transitions occurring in the phase diagrams, the band structures are presented in Fig. \ref{fig:pds_and_bands} for a particular value of $t_1$ and $t_\perp$, namely, $t_1 = t$ and $t_\perp = 0.5t$. The values of $\phi_l$ and $\phi_u$ are such that they lie along the four lines, namely, $L_1$, $L_2$, $L_3$ and $L_4$ in the phase diagrams depicted in Fig. \ref{fig:pds_bd}, and are denoted by $\eta_i \,(i=1,\dots,6)$, $\gamma_j \,(j=1,\dots,9)$, and $\chi_s \,(s=1,2,3)$. Along $L_1$, a topological phase transition occur between $C=+2$ and $C=-2$ corresponding to band-v2, while the transition between $C = +1$ and $C = -1$ occur along $L_2$ for both band-v2 and band-v1. These results have to be understood in conjunction with the corresponding band structures as shown in Figs. \ref{fig:band_eta1}-\ref{fig:band_eta3} and \ref{fig:band_eta4}-\ref{fig:band_eta6} respectively. The band structures corresponding to $\eta_1$ and $\eta_3$ points are identical and the Chern numbers corresponding to band-v2 are $+2$ and $-2$ respectively (Fig. \ref{fig:pd11_subfig}). At $\eta_2$, band-v2 and band-c2 touch each other at both the Dirac points (Fig. \ref{fig:band_eta2}), and hence there is a phase transition at $\eta_2$. However, band-v1 remains isolated from band-v2 at these $\eta$ points, and the Chern numbers are zero along $L_1$ as evident from its phase diagram (Fig. \ref{fig:pd21_subfig}). Further, the band structures corresponding to $\eta_4$ and $\eta_6$ have similar features, however in this case, $C$ has values $+1$ and $-1$ respectively corresponding to band-v2, while for band-v1, $C$ has the same magnitude, but are of opposite signs. At the phase transition occurring at $\eta_5$, band-v2 and band-v1 touch each other at the $\mathbf{K}$ point in the BZ (Fig. \ref{fig:band_eta5}), and hence for both the bands, a topological phase transition takes place at this point.
	
	Further, along $L_3$, again multiple phase transitions occur (see Figs. \ref{fig:pd12_subfig} and \ref{fig:pd22_subfig}), and the corresponding dispersions are shown in Fig. \ref{fig:band_gam1}-\ref{fig:band_gam9}. At $\gamma_1$, band-v2 and band-v1 show $C=+1$ and $C=-1$ respectively which drops to zero at $\gamma_2$ and hence the gap between those bands close at the $\mathbf{K}$ point as shown in Fig. \ref{fig:band_gam2}. At $\gamma_3$, the gap reopens, but the Chern numbers corresponding to these bands remain zero. The gap between band-v2 and band-c2 vanishes at $\gamma_4$ where again a phase transition takes place, since along the line connecting $\gamma_4$ and $\gamma_6$, the Chern number has a value $+2$. The band structure at an intermediate point, namely, $\gamma_5$ has been shown in Fig. \ref{fig:band_gam5}. Similarly,  phase transitions take place at $\gamma_6$ and $\gamma_8$, where the gaps vanish at the $\mathbf{K^\prime}$ point. At $\gamma_7$ and $\gamma_9$, $C$ assumes values zero and $+1$ respectively corresponding to band-v2. 
	It should be noted that band-v1 shows vanishing of the Chern number between $\gamma_2$ and $\gamma_8$ segments (see fig. \ref{fig:pd22_subfig}) and hence it never touches band-v2 which results in absence of any phase transition. 
	
	Now, we show the phase transitions between $C=-2$ and $C=-1$ phase along $L_4$. The corresponding band structures are shown in Fig. \ref{fig:band_chi1}-\ref{fig:band_chi3}. At $\chi_2$, band-v2 and band-v1 remain isolated from each other, however, they possess Chern numbers $C=-2$ and $C=0$ respectively. At $\chi_1$ and $\chi_3$, these two bands touch each other at the $\mathbf{K^\prime}$ and the $\mathbf{K}$ points in the BZ respectively, where topological phase transitions take place. Beyond $\chi_1$ and $\chi_3$, the gap reopens and both the bands possess non-trivial phases with $C=-1$. 
	
	Further, along the $\phi_u =-\phi_l$ line, a semi-metallic phase exists for all the bands. In the vicinity of $\phi_u = \phi_l$, only the phase diagrams of band-v1 show trivial regions with $C=0$, however, those for band-v2 demonstrate non-trivial phases either with $C=+2$ or $C=-2$.

\begin{figure}[h]
		\captionsetup[subfigure]{labelformat=nocaption}
		\centering
		\includegraphics[width=\linewidth]{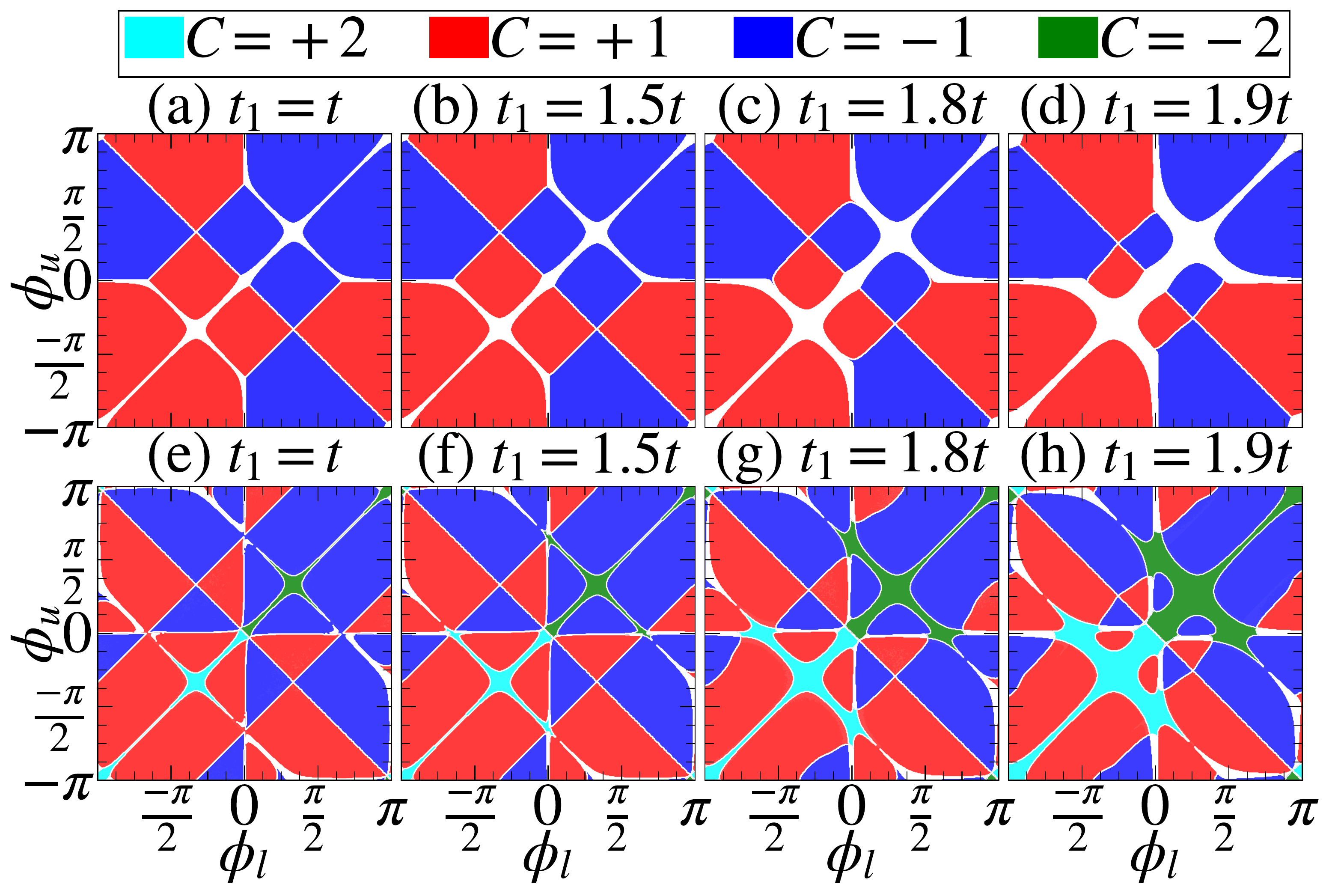}

		\begin{subfigure}[b]{0\textwidth}
			\subcaption{}\label{fig:pd_phi_t1_1_tcp_0.1_bd_1}
		\end{subfigure}
		\begin{subfigure}[b]{0\textwidth}
			\subcaption{}\label{fig:pd_phi_t1_1.5_tcp_0.1_bd_1}
		\end{subfigure}
		\begin{subfigure}[b]{0\textwidth}
			\subcaption{}\label{fig:pd_phi_t1_1.8_tcp_0.1_bd_1}
		\end{subfigure}
		\begin{subfigure}[b]{0\textwidth}
			\subcaption{}\label{fig:pd_phi_t1_1.9_tcp_0.1_bd_1}
		\end{subfigure}
		
		\begin{subfigure}[b]{0\textwidth}
			\subcaption{}\label{fig:pd_phi_t1_1_tcp_0.1_bd_2}
		\end{subfigure}
		\begin{subfigure}[b]{0\textwidth}
			\subcaption{}\label{fig:pd_phi_t1_1.5_tcp_0.1_bd_2}
		\end{subfigure}
		\begin{subfigure}[b]{0\textwidth}
			\subcaption{}\label{fig:pd_phi_t1_1.8_tcp_0.1_bd_2}
		\end{subfigure}
		\begin{subfigure}[b]{0\textwidth}
			\subcaption{}\label{fig:pd_phi_t1_1.9_tcp_0.1_bd_2}
		\end{subfigure}
		
		\caption{\raggedright The phase diagrams corresponding to band-v1 are shown in (a)-(d), while those for band-v2 are shown in (e)-(h). The value of $t_\perp$ is kept fixed at $0.1t$. The values of $t_1$ are such that $t_1=t$ in (a) and (e), $t_1=1.5t$ in (b) and (f), $t_1=1.8t$ in (c) and (g), and $t_1=1.9t$ in (d) and (h). The white regions denote trivial phases with zero Chern numbers, while the colored regions indicate the non-trivial phases with the non-zero Chern numbers. The values are indicated at the top of the figure.}
		\label{fig:pd_phiphi_tcp_0.1_bd_1and2}
	\end{figure}

	Moreover, in order to see the effects of $t_\perp$ on the topological phases, we have shown the phase diagrams corresponding to band-v1 and band-v2 in Figs. \ref{fig:pd_phi_t1_1_tcp_0.1_bd_1}-\ref{fig:pd_phi_t1_1.9_tcp_0.1_bd_1} and \ref{fig:pd_phi_t1_1_tcp_0.1_bd_2}-\ref{fig:pd_phi_t1_1.9_tcp_0.1_bd_2} respectively. It is evident that the areas of Chern insulating regions are enhanced corresponding to lower values of $t_\perp$. Also, the shape of the topological regions are different from those for the $t_\perp = 0.5t$ case. Further, the areas of  $C = \pm 2$ regions in the phase diagram corresponding to band-v2 are mostly spanned by $C = \pm 1$ regions. However, the feature that remains unaltered is the trivial phase along $\phi_u = \pm\phi_l$ lines for band-v1 and $\phi_u = \phi_l$ line for band-v2. For both $t_\perp=0.5t$ and $t_\perp = 0.1t$, the Chern insulating regions gradually shrink with the increase in the value of $t_1$ and finally vanish at the semi-Dirac limit, namely, $t_1 = 2t$.

\section{Edge states}\label{sec:edge_states}
	
	\begin{figure}[h]
		\centering
		\begin{subfigure}[b]{0.49\linewidth}
			\includegraphics[width=\textwidth]{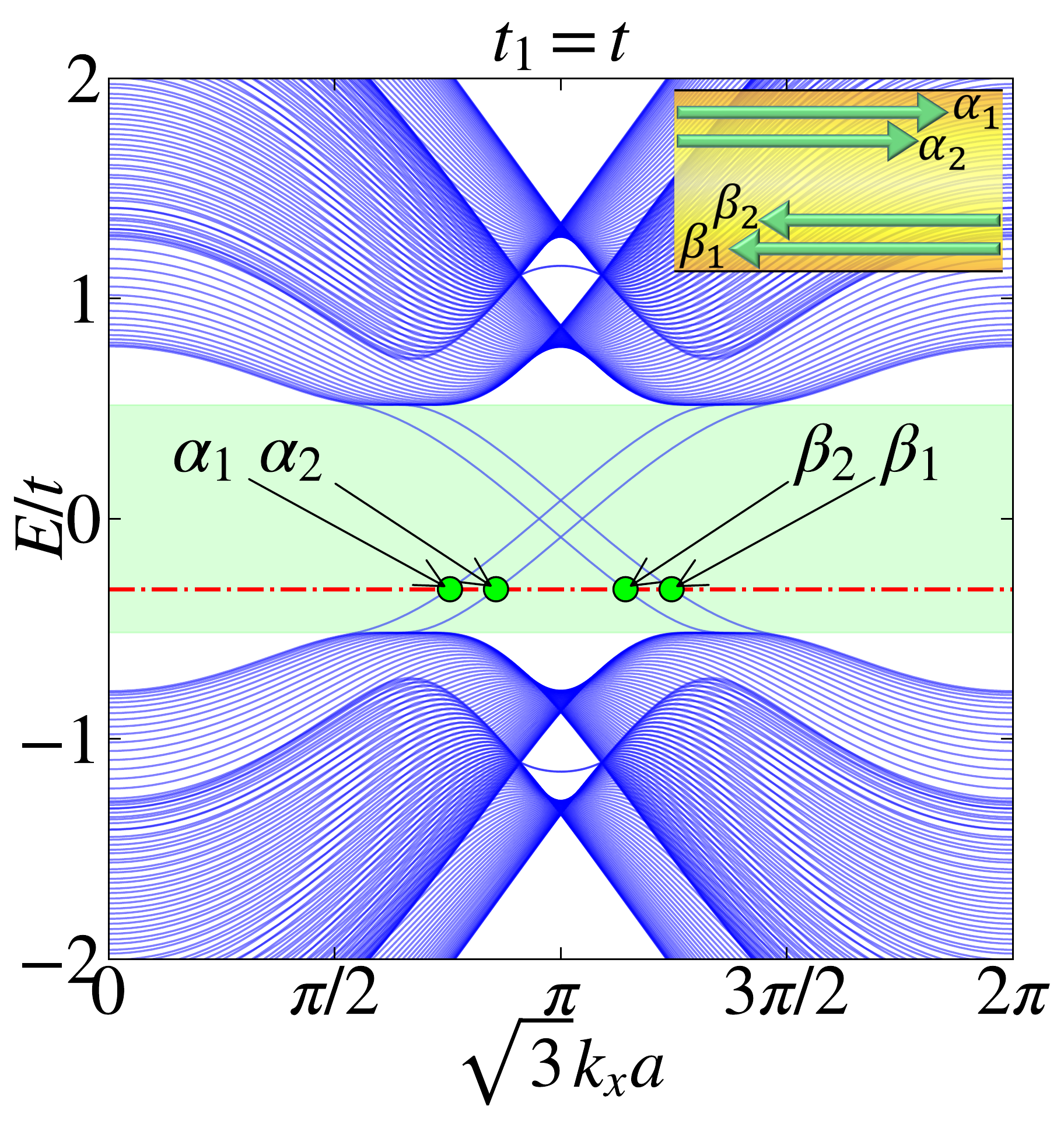}
			\subcaption{}\label{fig:ES_t1_1_tcp_0.5}
		\end{subfigure}
	\begin{subfigure}[b]{0.49\linewidth}
		\includegraphics[width=\textwidth]{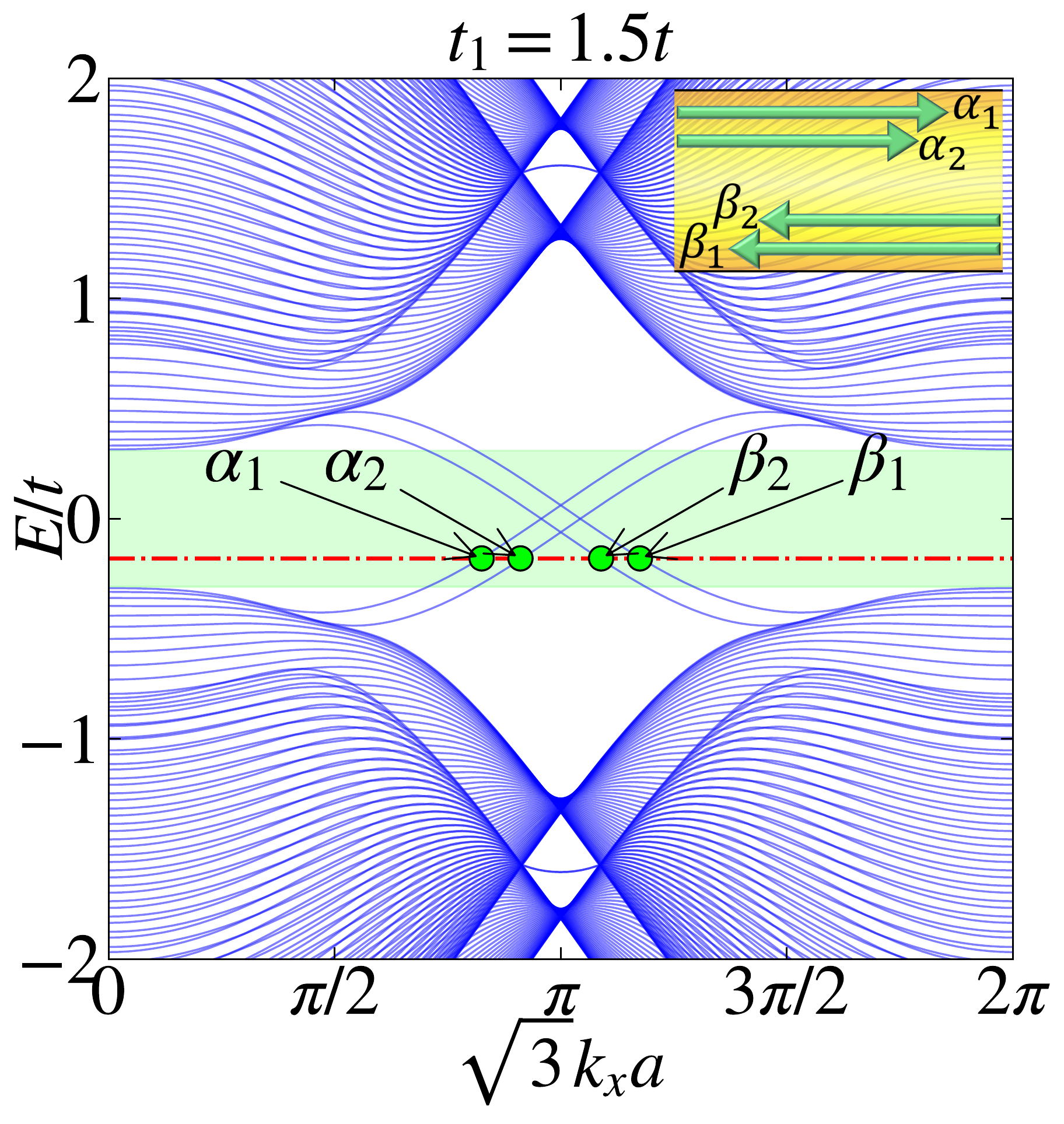}
		\subcaption{}\label{fig:ES_t1_1.5_tcp_0.5}
	\end{subfigure}
	\begin{subfigure}[b]{0.49\linewidth}
		\includegraphics[width=\textwidth]{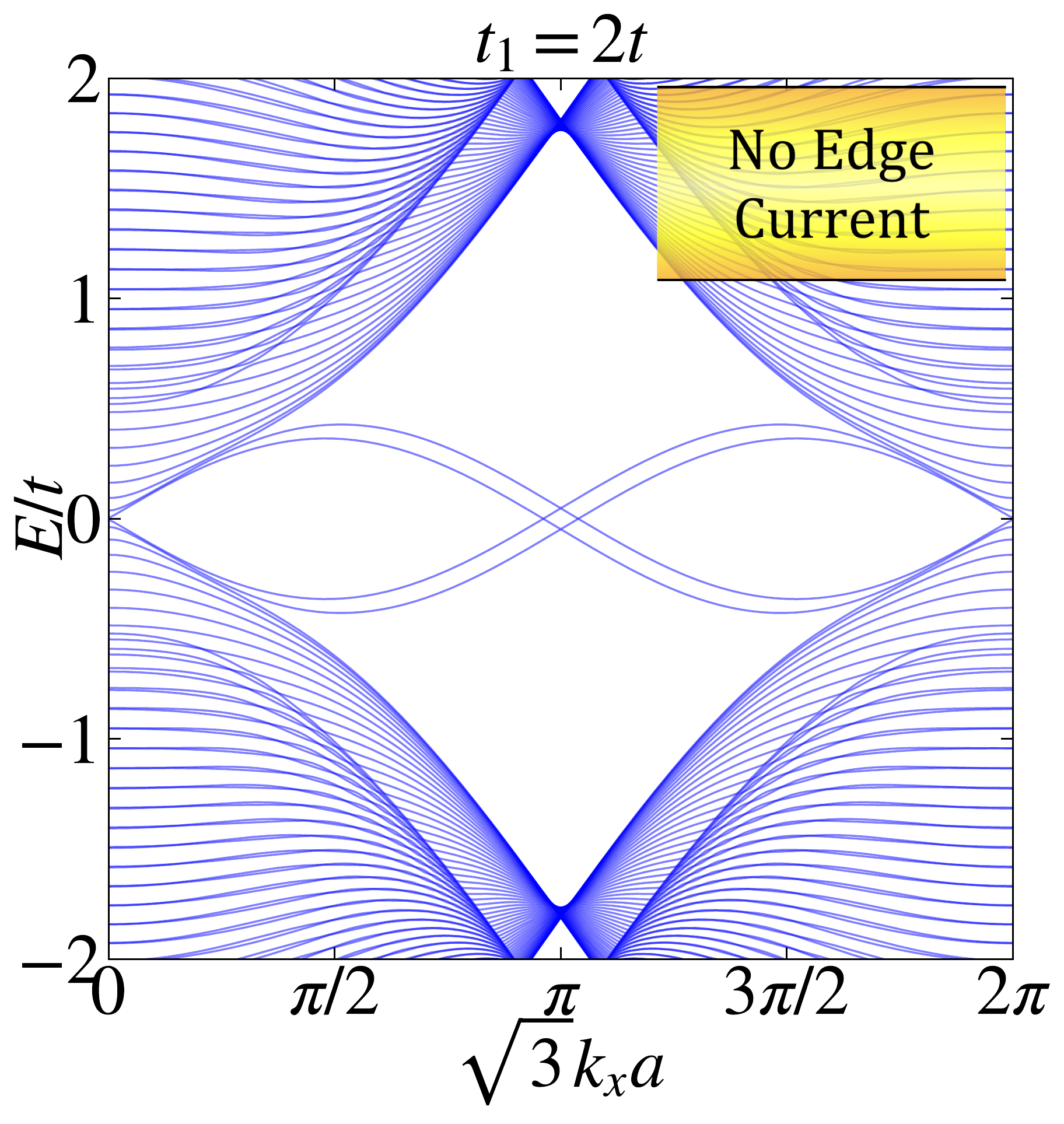}
		\subcaption{}\label{fig:ES_t1_2.0_tcp_0.5}
	\end{subfigure}
	\begin{subfigure}[b]{0.49\linewidth}
		\includegraphics[width=\textwidth]{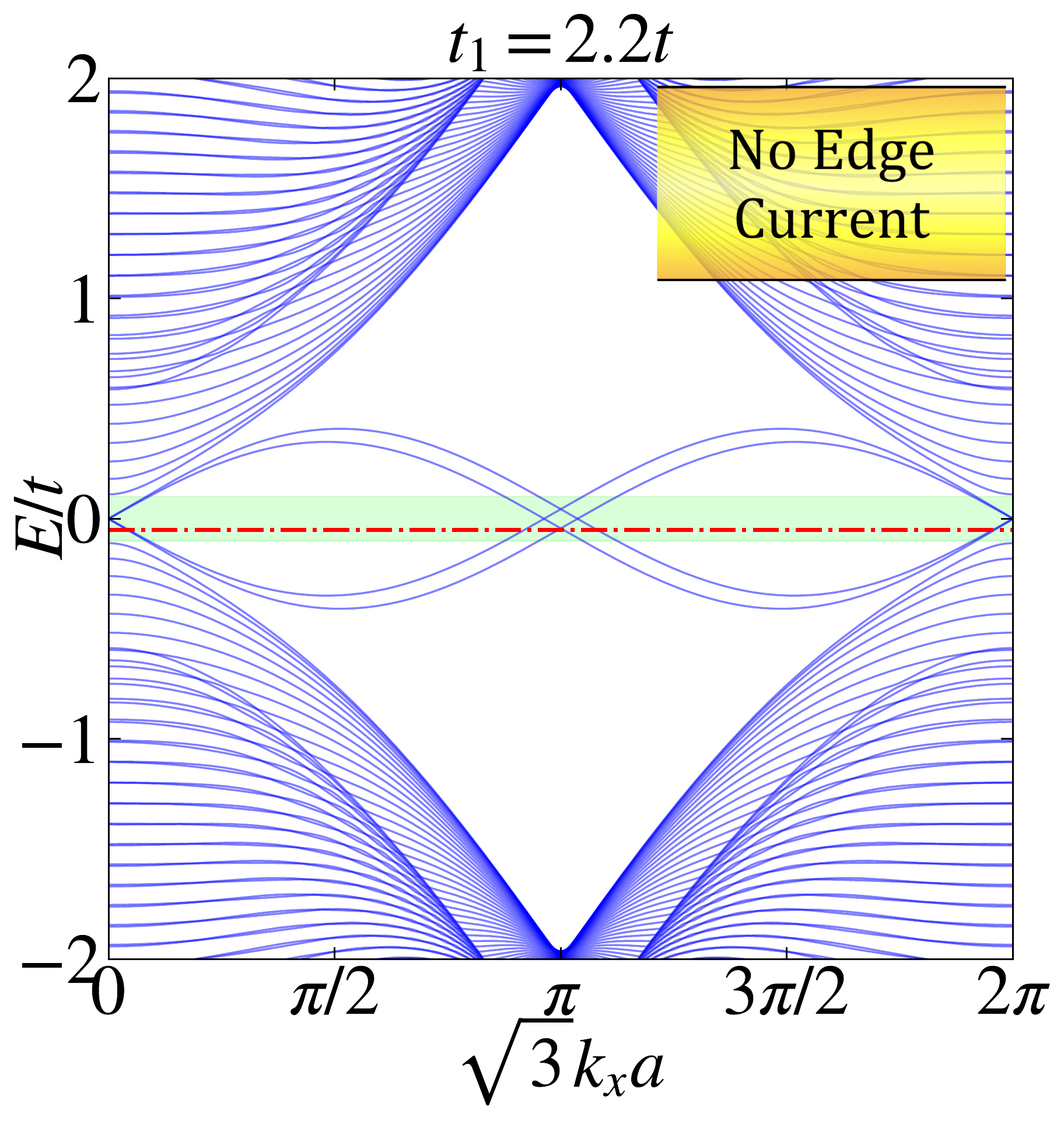}
		\subcaption{}\label{fig:ES_t1_2.2_tcp_0.5}
	\end{subfigure}
		
		\caption{\raggedright The edge state spectra are shown for (a) $t_1=t$, (b) $t_1=1.5t$, (c) $t_1=2t$, and (d) $t_1=2.2t$. The green shaded regions represent the bulk gap in (a), (b) and (d) (np bulk gap in (c)). The Fermi levels ($E_F$) are denoted by the red dashed lines, which are shown to be present in the bulk gap. $E_F$ intersects the edge modes at the points denoted by the green dots as shown in (a) and (b). For these, the edge currents are shown by the green arrows in the yellow panels located at the top right corner, which represent parts of the semi-infinite ribbon.}
		\label{fig:ES}
	\end{figure}
	
	To show the existence (and their vanishing) of the edge modes, in this section, we show the band structure of the system for semi-infinite nanoribbon. The ribbon has a finite width along the $y$-direction, while it is infinite along the $x$-direction \cite{nakada1996, sticlet2012}. Further, we label the sites along the $y$-direction as A$_1^l$, B$_1^l$, A$_2^l$, B$_2^l$, .... A$_N^l$, B$_N^l$, A$_1^u$, B$_1^u$, A$_2^u$, B$_2^u$, .... A$_N^u$, B$_N^u$. Since, the periodicity along the $x$-direction remains preserved, we can Fourier transform the operators along that direction. This results in set of four coupled equations as shown below. 
	
	\begin{equation}\label{eq:edge1}
		\begin{aligned}
			E_{k} a_{k, n}^u =&-\left[ t\left\{1+ e^{(-1)^n ik} \right\}b_{k, n}^u + t_1 b_{k, n-1}^u \right]\\ &-2t_2\left[ \cos(k + \phi)a_{k, n}^u + e^{(-1)^n\frac{ik}{2}}\times \right.\\ &\left. \cos\left( \frac{k}{2} - \phi\right) \{a_{k, n-1}^u + a_{k, n+1}^u \} \right]
		\end{aligned}
	\end{equation}
	
		\begin{equation}\label{eq:edge2}
			\begin{aligned}
				&E_{k} b_{k, n}^u =-\left[ t\left\{ 1+ e^{(-1)^{n+1} ik} \right\}a_{k, n}^u + t_1 a_{k, n+1}^u \right]\\ &-2t_2\left[  e^{(-1)^{n+1} \frac{ik}{2}}\times  \cos\left( \frac{k}{2} + \phi\right)\{a_{k, n-1}^u + a_{k, n+1}^u\} + \right.\\ & \left. \cos(k - \phi)b_{k, n}^u \right] + t_\perp\left[ \xi_1 e^{-ik} + \xi_2 \right]a_n^l
			\end{aligned}
		\end{equation}  
	\begin{equation}\label{eq:edge3}
	\begin{aligned}
		&E_{k} a_{k, n}^l =-\left[ t\left\{1+ e^{(-1)^n ik} \right\}b_{k, n}^l + t_1 b_{k, n-1}^l \right]\\ &-2t_2\left[ e^{(-1)^n\frac{ik}{2}}\times  \cos\left( \frac{k}{2} - \phi\right) \{a_{k, n-1}^l + a_{k, n+1}^l \}\right.  \\ 
		&\left. \cos(k + \phi)a_{k, n}^l + \right] + t_\perp\left[ \xi_1 e^{ik} + \xi_2 \right]b_n^u
	\end{aligned}
	\end{equation}
		\begin{equation}\label{eq:edge4}
			\begin{aligned}
				E_{k} b_{k, n}^l =&-\left[ t\left\{ 1+ e^{(-1)^{n+1} ik} \right\}a_{k, n}^l + t_1 a_{k, n+1}^l \right]\\ &-2t_2\left[ \cos(k - \phi)b_{k, n}^l + e^{(-1)^{n+1} \frac{ik}{2}}\times \right.\\ & \left. \cos\left( \frac{k}{2} + \phi\right)\{a_{k, n-1}^l + a_{k, n+1}^l \} \right]
			\end{aligned}
		\end{equation}
	where $a_{k, n}^{l,u}$ and $b_{k, n}^{l,u}$ are the amplitudes of the wave functions corresponding to the sublattices A and B respectively. The superscripts $l$ and $u$ refers to lower and upper layers respectively. Here $k = \sqrt{3}k_xa_0$ is the dimensionless momentum and $n$ denotes the site index which assumes integer values in the range $[1:N]$ with $N$ being the total number of unit cells along the $y$-direction. We chose $N$ as 128, which gives the width to be $79\sqrt{3}a_0$. In Eqs. \ref{eq:edge2} and \ref{eq:edge3}, $\xi_1$ and $\xi_2$ denote quantities that depend on the site index, $n$ via $\xi_1 = [ 1-(-1)^n ]/2$ and $\xi_2 = [ 1+(-1)^n ]/2$ respectively. 
	
	By solving Eqs. \ref{eq:edge1}, \ref{eq:edge2}, \ref{eq:edge3}, and \ref{eq:edge4}, we obtain the band structure of the nanoribbon for various values of $t_1$ as presented in Fig. \ref{fig:ES}. It can be noticed that a pair of edge modes from the valence bands (band-v2) traverse the Fermi level, $E_F$ (shown via the red dashed line) and merge with the conduction bands (band-c2) and another pair crosses the Fermi level in the opposite direction. Such crossing of the edge modes leads to a quantized Hall conductivity should the Fermi level lies in the bulk gap. $E_F$ intersects the edge modes (see Figs. \ref{fig:ES_t1_1_tcp_0.5} and \ref{fig:ES_t1_1.5_tcp_0.5}) at four points (marked by the green dots), whose corresponding edge currents are shown by the green arrows in the yellow panels located at the top right corner of the plots. The yellow panels represent a part of the semi-infinite ribbon. Since the velocity of electrons are proportional to the slope of the band structure, that is,  $\partial E / \partial k$, there exists a pair of edge currents at each edge that moves in the same direction. However, such pairs of currents propagate in opposite directions at the two edges of the ribbon. Hence, these modes are called chiral edge modes.
	
	It should be noted that because of a pair of chiral edge modes, we should obtain the Hall conductivity quantized with a plateau at $2e^2/h$, with the factor `2' arise due to doubling of the number of chiral edge modes \cite{hatsugai1988}. Such chiral edge modes exist as long as the value of $t_1$ remains lesser than $2t$. Since the bulk gap vanishes at $t_1 = 2t$ (see Fig. \ref{fig:ES_t1_2.0_tcp_0.5}), the edge current vanishes. For $t_1>2t$, the edge modes get detached from the bulk bands as shown in Fig. \ref{fig:ES_t1_2.2_tcp_0.5} for $t_1 = 2.2t$, thereby resulting in a zero edge current. These results are consistent with the corresponding Chern numbers obtained in the phase diagram. For example, we observe the non-zero edge currents for $t_1<2t$ and the corresponding Chern number is found to be $C = |2|$. For $t_1>2t$, the Chern numbers vanish, and so the edge currents. The figures presented here are for $t_\perp = 0.5t$. For $t_\perp = 0.1t$, we observe similar features in the spectrum, except that the bulk gaps get reduced. We have skipped the discussion of the latter for brevity.
	
	\section{Hall conductivity} \label{sec:hall_conductivity}
	
	\begin{figure}[h]
		\centering
		\begin{subfigure}[b]{0.8\linewidth}
			\includegraphics[width=\textwidth]{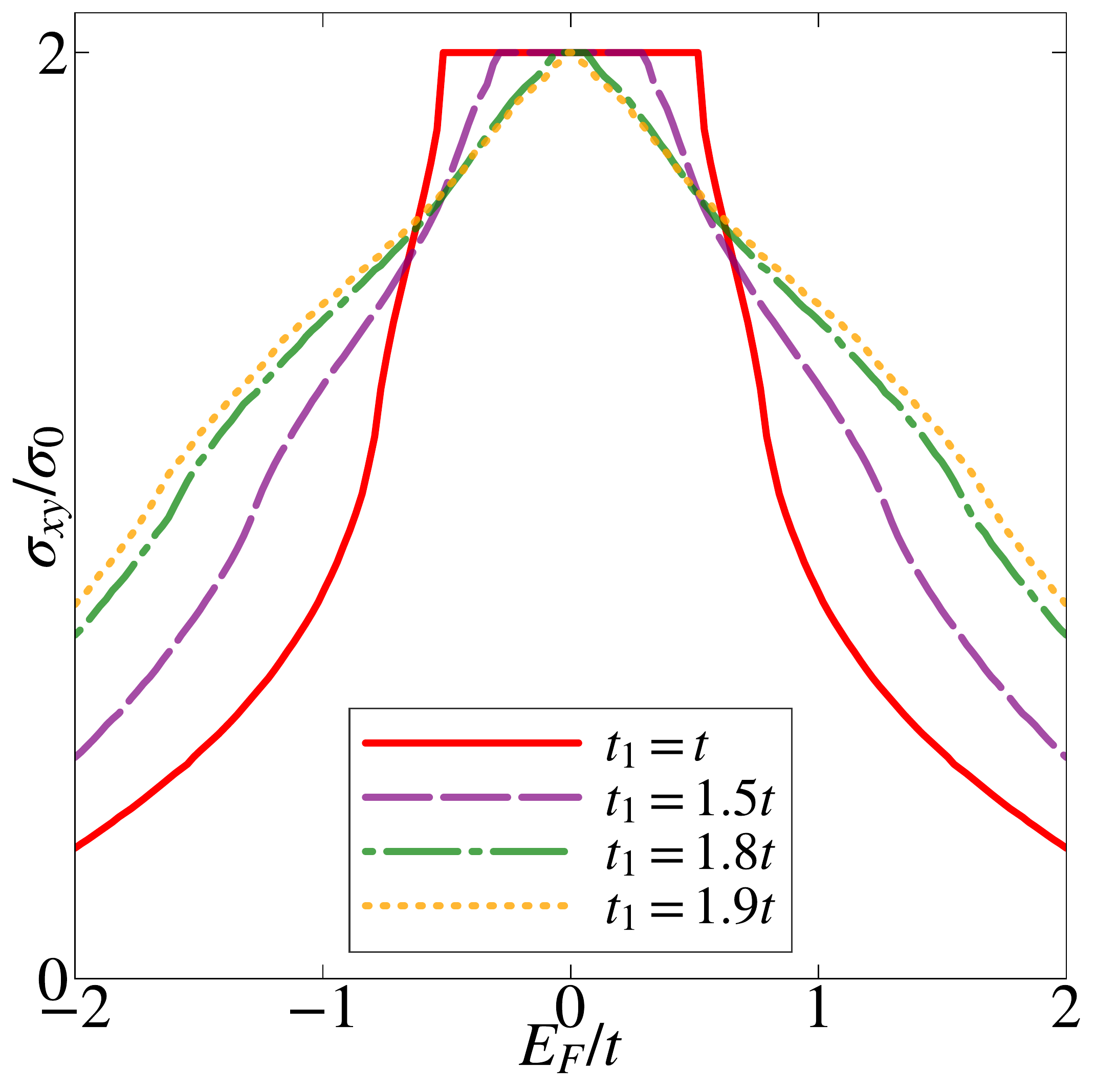}
			\subcaption{}\label{fig:hall_tc_0.5t}
		\end{subfigure}\\
		\begin{subfigure}[b]{0.8\linewidth}
			\includegraphics[width=\textwidth]{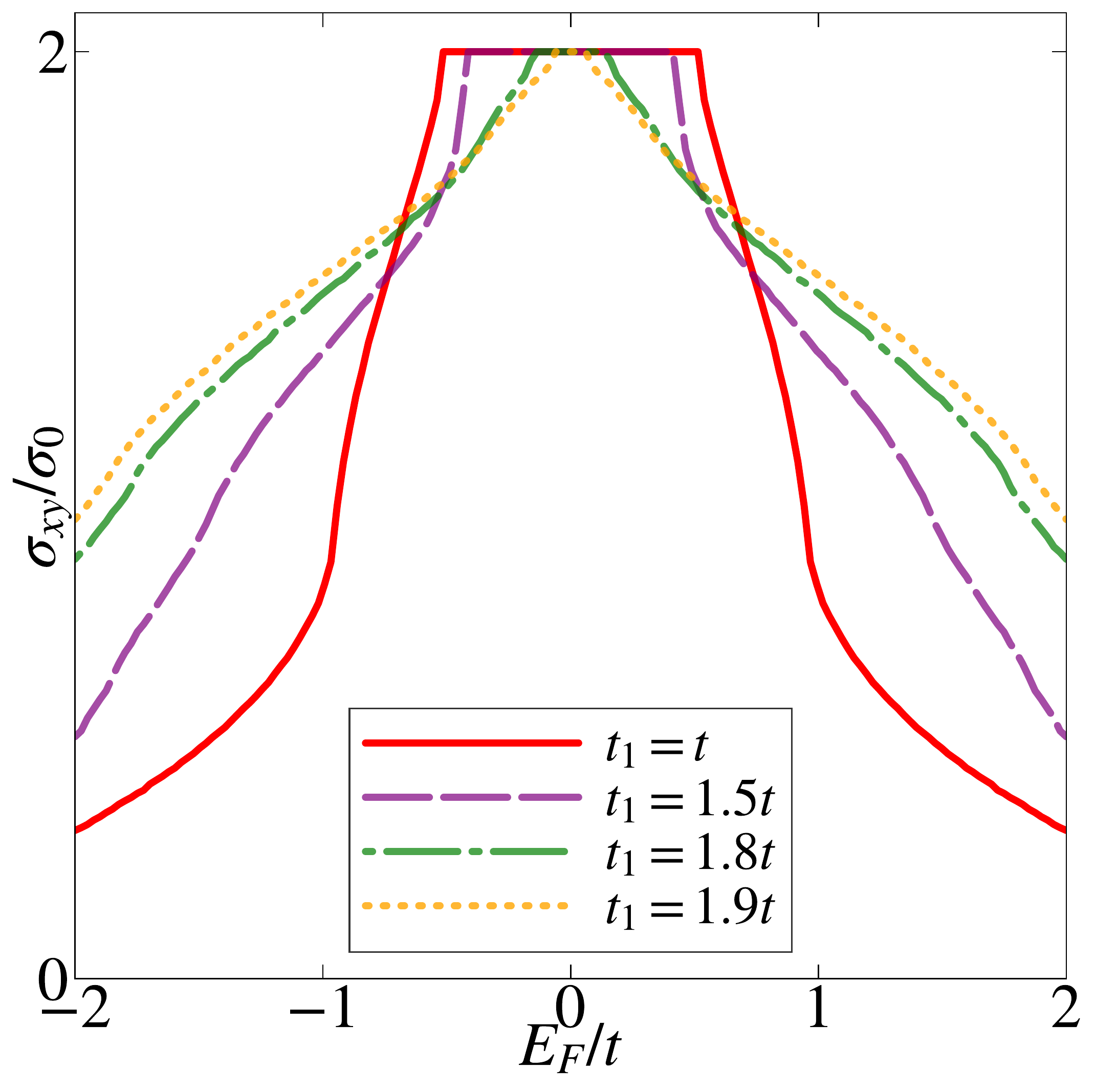}
			\subcaption{}\label{fig:hall_tc_0.1t}
		\end{subfigure}
		
		\caption{\raggedright The anomalous Hall conductivities are shown as a function of $E_F$ for various values of $t_1$ in (a) and (b) for $t_\perp=0.5t$ and $t_\perp=0.1t$ respectively. The plateau width decreases as $t_1$ deviates from $t$.}
		\label{fig:hall_cond}
	\end{figure}
	
	In this section, we calculate the anomalous Hall conductivity as function of the Fermi energy $E_F$. The prerequisite is the computations of the Berry curvature using Eq. \ref{eq:berry_curv} and hence use the following formula to calculate the anomalous Hall conductivity ($\sigma_{xy}$) \cite{hall1,hall2}, namely,
	\begin{equation}\label{eq:Hall_cond}
		\sigma_{xy} = \frac{\sigma_0}{2\pi} \sum_{\lambda} \int \frac{\mathrm{d}k_x \mathrm{d}k_y}{(2\pi)^2} f\left(E^\lambda_{k_x, k_y} \right) \Omega(k_x, k_y)
	\end{equation}
	where $f(E^\lambda_{k_x, k_y}) = \left[ 1+\exp{\{(E^\lambda_{k_x, k_y}-E_F)/K_BT\}} \right]^{-1}$ is the Fermi-Dirac distribution function at an energy $E^\lambda_{k_x, k_y}$. Here $E_F$ refers to the Fermi energy and $T$ is the absolute temperature. The energy is denoted by $E^\lambda_{k_x, k_y}$ with $\lambda$ being the band index. The constant term $\sigma_0$ is equal to $e^2/h$ which sets the scale for $\sigma_{xy}$. Now, we compute $\sigma_{xy}$ numerically as a function of $E_F$ at zero temperature ($T=0$) for various values of $t_1$ as shown in Fig. \ref{fig:hall_cond}.
	
	As can be seen from Fig. \ref{fig:hall_tc_0.5t}, when the Fermi energy, $E_F$ lies in the bulk gap, $\sigma_{xy}$ becomes quantized at a value $2\sigma_0$. The width of the plateau is equal to the width of the bulk gap in the dispersion spectrum of Fig. \ref{fig:band1}. As soon as $E_F$ intersects the bands (either both the conduction or both the valence bands), $\sigma_{xy}$ starts to decrease since the integral is performed over the occupied states. This also results in diminishing of the plateau width with increase in the value of $t_1$. This happens because the energy gap between the band-c2 and the band-v2 shrinks. The plateau and the Hall conductivity vanish completely at $t_1=2t$, where the spectrum becomes gapless. For the hopping asymmetry engineered beyond the semi-Dirac limit, that is, $t_1>2t$, the bands become gapped again, however, the Hall conductivity remains zero. These results are consistent with their corresponding information coming from the Chern numbers. The Hall plateaus are observed as long as the system remains a Chern insulator, that is, for $t_1<2t$. Further, the factor `2' in ($2\sigma_0$) denotes the value of Chern number (and also the edge modes) which vanishes for $t_1>2t$.
	
	We have also presented the Hall conductivity for a smaller value of $t_\perp$, namely, $t_\perp=0.1t$ in Fig. \ref{fig:hall_tc_0.1t}. In this case, the plateau widths corresponding to different values of $t_1$ are larger as compared to that for the $t_\perp = 0.5t$ case, since the corresponding band gaps are enhanced as shown in Fig. \ref{fig:band2}. However, similar to the previous case, the plateau width decreases with the increase of $t_1$, which finally vanishes at $t_1 = 2t$ and beyond. Thus, a topological phase transition takes place across the gap closing point at the semi-Dirac limit, namely, $t_1 = 2t$.
	
	\section{Conclusion}\label{sec:conclusion}
	We have investigated the topological properties for a band engineered bilayer Haldane. By tuning one of the three NN hopping amplitudes, the band extrema, which were located at the  $\mathbf{K}$ and the $\mathbf{K^\prime}$ points for the Dirac case, migrate towards each other and finally merge at an intermediate $\mathbf{M}$ point in the BZ in the semi-Dirac limit, that is, at $t_1=2t$. We have calculated the Chern numbers for various values of $t_1$ and plotted them in the $\phi_u$-$\phi_l$ plane which demonstrates that the higher Chern numbers ($C=\pm2$) are associated only with band-v2. However, the Chern numbers corresponding to both both bands vanish, that is, there are topological phase transitions, where the Chern numbers discontinuously change from $C=\pm2$ to $C=0$ and $C = \pm1$ to $C = 0$, across the semi-Dirac point $t_1 = 2t$. Also, there are multiple phase transitions in the phase diagram, such as, $+2\rightarrow-2$, $+1\rightarrow-1$, $\pm2\rightarrow\pm 1$ and $\pm2 \rightarrow 0 \rightarrow 1$. These phase transitions are confirmed by the opening and closing of the energy gaps (semi-metallic phase) in the dispersion spectrum. Further, we have also computed the band structure of a nanoribbon, where we observe a pair of chiral edge modes along the edges of the ribbon exist as long as $t_1$ remains lesser than $2t$. Also for the anomalous Hall conductivity, the width of the quantized plateau at $2\sigma_0$ gradually decreases with increase in $t_1$, which finally vanishes at $t_1=2t$. Thus, a bilayer Haldane model, similar to its monolayer analogue, exhibits a topological phase transition at the semi-Dirac point. However, here we have larger values of the Chern number and doubling of the edge modes at the edges of the bilayer nanoribbon. Further, the phase transitions are supported by the vanishing of Chern number, chiral edge modes and the anomalous Hall conductivity.


\begin{thebibliography}{8}
\bibitem{phiphi_pd}
However, for obtaining the Chern number phase diagrams see Figs. \ref{fig:pd_phiphi_tcp_0.5_bd_1}, \ref{fig:pd_phiphi_tcp_0.5_bd_2}, and \ref{fig:pd_phiphi_tcp_0.1_bd_1and2}, $\phi_l$ and $\phi_u$ are varied in the range $[-\pi:+\pi]$.
\end{thebibliography}
\end{document}